\documentclass[a4paper,11pt]{article}
\usepackage{jheppub} 

\usepackage{subcaption} %
\usepackage{booktabs}
\usepackage{multirow}  %
\usepackage{siunitx}  %
\usepackage{xcolor}
\usepackage{amsfonts} 

\graphicspath{{figures/}}

\usepackage{cleveref}
\crefname{figure}{figure}{figures}     
\Crefname{figure}{Figure}{Figures}     
\crefrangemultiformat{figure}{figures~#3#1#4--#5#2#6}{}{}{} 
\crefmultiformat{figure}{figures~#2#1#3}{ and~#2#1#3}{, #2#1#3}{ and~#2#1#3} 

\title{Low-energy interactions between doubly charmed baryons and Goldstone bosons from lattice QCD}

\author[a,b,d]{Jing-Yu Yi}
\emailAdd{yijingyu@hnu.edu.cn}
\author[c,a]{Ze-Rui Liang}
\emailAdd{liangzr@hebtu.edu.cn}
\author[d,e,\ast]{Liuming Liu\note[$\ast$]{Corresponding author}}
\emailAdd{liuming@impcas.ac.cn}
\author[a,b,\ast]{De-Liang Yao}
\emailAdd{yaodeliang@hnu.edu.cn}

\affiliation[a]{School of Physics and Electronics, Hunan University, Changsha 410082, China}
\affiliation[b]{Hunan Provincial Key Laboratory of High-Energy Scale Physics and Applications, Hunan University, Changsha 410082, China}
\affiliation[c]{College of Physics and Hebei Key Laboratory of Photophysics Research and Application, Hebei Normal University, Shijiazhuang, Hebei 050024, China}
\affiliation[d]{Institute of Modern Physics, Chinese Academy of Sciences,
 Lanzhou 730000, China}
\affiliation[e]{University of Chinese Academy of Sciences, School of Physical Sciences,
Beijing 100049, China}

\abstract{
We perform a lattice QCD calculation of the $S$-wave interactions between the ground-state spin-$1/2$ doubly charmed baryons and Goldstone bosons. The lattice QCD simulations are carried out on four $2+1$ flavor Wilson-Clover ensembles generated by the CLQCD collaboration, with a lattice spacing $a=0.07746$ fm and two different pion masses, $M_\pi \sim 210$ and $\sim 300~\mathrm{MeV}$. Energy levels are extracted for four single channels, $\Omega_{cc}\bar{K}^{(-2,1/2)}$, $\Xi_{cc}K^{(1,1)}$, $\Xi_{cc}K^{(1,0)}$, and $\Xi_{cc}\pi^{(0,3/2)}$, where the superscripts $(S,I)$ denote strangeness $S$ and isospin $I$. Our results indicate that the $\Xi_{cc}K^{(1,0)}$ channel is attractive, exhibiting negative energy shifts relative to the non-interacting two-hadron thresholds, while the other three channels are repulsive. Using Lüscher’s finite-volume formula, we extract the near-threshold phase shifts and determine the $S$-wave scattering lengths. Furthermore, a virtual state pole is found in the $\Xi_{cc}K^{(1,0)}$ scattering amplitude. These results provide {\it ab initio} input to enable high-precision studies of the properties and spectroscopy of doubly heavy baryons.
}

\allowdisplaybreaks[4]

\begin{document}
\maketitle

\section{Introduction}

The study of doubly charmed baryons (DCBs), predicted by the SU(4) version of the traditional quark model~\cite{Gell-Mann:1964ewy}, plays an essential role not only in completing the charmed hadron spectroscopy~\cite{ParticleDataGroup:2024cfk} but also in probing the non-perturbative dynamics of low-energy quantum chromodynamics (QCD) in the presence of two heavy quarks. These states, consisting of two charm quarks and one light quark, serve as ideal systems for investigating heavy quark symmetry, chiral dynamics, and strong interaction mechanisms that are not easily accessible in singly-heavy or light hadrons; see, e.g., refs.~\cite{Yao:2020bxx,Cheng:2021qpd,Meng:2022ozq,Chen:2022asf,Crede:2024hur} for recent reviews.

In the past decade, significant experimental progress has been made in the investigation of DCBs; however, many questions remain open. The first observation of $\Xi_{cc}^{++}$ was reported by the LHCb collaboration~\cite{LHCb:2017iph} in 2017 through the decay mode $\Lambda_c^+ K^-\pi^+\pi^+$, with a mass of approximately 3621 MeV and a statistical significance exceeding 12$\sigma$. Note that the discovery potential of $\Xi_{cc}^{++}$ in the $\Lambda_c^+ K^-\pi^+\pi^+$ mode was suggested by the theoretical work~\cite{Yu:2017zst}. This finding was subsequently supported by observations in other decay channels such as $\Xi_{cc}^{++} \to \Xi_c^{(\prime)+}\pi^+$~\cite{LHCb:2018pcs, LHCb:2022rpd} and $\Xi_{cc}^{++} \to \Xi_c^0\pi^+\pi^+$~\cite{LHCb:2025shu}. More precise measurements followed, yielding improved mass and lifetime estimates~\cite{LHCb:2019epo, LHCb:2018zpl}, as well as production cross-section data in proton-proton collisions at 13 TeV~\cite{LHCb:2019qed}. These results are broadly consistent with predictions from quark models~\cite{Ebert:2002ig, Roberts:2007ni}, as well as with those from QCD sum rules~\cite{Kiselev:2001fw, Zhang:2008rt, Wang:2010hs} and earlier lattice QCD calculations~\cite{Lewis:2001iz, Flynn:2003vz, Liu:2009jc}. In contrast, its isospin partner $\Xi_{cc}^+$ and the strange cousin $\Omega_{cc}^+$ in the same SU(3) triplet have not yet been firmly observed. An earlier claim of $\Xi_{cc}^+$ observation by the SELEX collaboration~\cite{SELEX:2002wqn, SELEX:2004lln}, with a mass near 3520 MeV, was not corroborated by subsequent experiments such as FOCUS~\cite{Ratti:2003ez}, BaBar~\cite{BaBar:2006bab}, Belle~\cite{Belle:2006edu}, or LHCb~\cite{LHCb:2013hvt, LHCb:2019gqy}. Moreover, the SELEX mass value is in tension with modern theoretical predictions (e.g., Refs.~\cite{Yao:2018ifh,Ortiz-Pacheco:2023kjn,Li:2025rtp}). Recent searches conducted by LHCb for decay signatures of $\Xi_{cc}^+$ and $\Omega_{cc}^+$ have not yielded significant signals~\cite{LHCb:2021rkb, LHCb:2021eaf}, indicating that their production and decay mechanisms are not yet well understood. 

Complementary to experimental measurements, lattice QCD provides a first-principle approach to explore the properties of DCBs directly from QCD. Extensive and high-precision calculations of the DCB spectrum have been performed using a variety of lattice formulations~\cite{Lewis:2001iz, Flynn:2003vz, Na:2007pv, Na:2008hz, Liu:2009jc, Basak:2012py, Briceno:2012wt, Durr:2012dw, PACS-CS:2013vie, Brown:2014ena, Padmanath:2015jea, Perez-Rubio:2015zqb, Alexandrou:2017xwd, Mathur:2018rwu, Bahtiyar:2020uuj, Alexandrou:2023dlu}, including simulations with $N_f=2+1$ and $N_f=2+1+1$ dynamical fermions. These state-of-the-art calculations employ strategies such as computations directly at the physical pion mass~\cite{PACS-CS:2013vie, Alexandrou:2017xwd, Alexandrou:2023dlu}, as well as those at unphysical pion masses followed by controlled continuum extrapolations, e.g., ref.~\cite{Mathur:2018rwu}. The most recent lattice QCD result~\cite{Alexandrou:2023dlu}, $M_{\Xi_{cc}}=3.634(51)$ GeV, is in excellent agreement with the LHCb measurement of $\Xi_{cc}^{++}$~\cite{LHCb:2017iph} and disfavours the earlier SELEX value~\cite{SELEX:2002wqn, SELEX:2004lln}. Beyond the mass spectrum, lattice QCD has also been applied to explore a range of other properties, including electromagnetic form factors~\cite{Can:2013zpa,Can:2013tna,Can:2021ehb,Bahtiyar:2022nqw}, radiative transitions~\cite{Bahtiyar:2018vub}, and thermal behavior~\cite{Aarts:2023nax}, which provide further insights into the structural and dynamic characteristics of the DCBs.

Despite these advances, a lattice QCD study of the interactions between DCBs and Goldstone bosons is still missing. Nevertheless, previous efforts have been made by imposing baryon chiral perturbation theory (BChPT)~\cite{Gasser:1983yg,Gasser:1984gg,Gasser:1987rb}, the effective field theory~\cite{Weinberg:1978kz} of QCD at low energies. In ref.~\cite{Meng:2018zbl}, the $S$-wave scattering lengths of the DCBs $B_{cc}\in\{\Xi_{cc}^{++},\Xi_{cc}^{+},\Omega_{cc}^{+}\}$ and the Goldstone bosons $\phi\in\{\pi^0,\pi^\pm,K^0,K^\pm,\bar{K}^0,\eta\}$ are derived within the heavy baryon (HB) formalism of BChPT. The non-relativistic calculation~\cite{Meng:2018zbl} is upgraded to a relativistic one~\cite{Liang:2023scp} by using the extended-on-mass-shell (EOMS) renormalization scheme~\cite{Fuchs:2003qc}, where the $S$- and $P$-wave scattering lengths, together with the $S$-wave phase shifts, are systematically predicted. The chiral amplitudes have been utilized to investigate the dynamically generated exotic DCB baryons in Refs.~\cite{Guo:2017vcf,Yan:2018zdt}. By using the $\mathcal{O}(p^2)$ chiral potentials and incorporating the $P$-wave excitations between the two charm quarks, two quasistable states in the spectrum of negative-parity DCBs are found in ref.~\cite{Yan:2018zdt}.  However, the aforementioned BChPT predictions strongly rely on the values of the low-energy constants (LECs), stemming from the $B_{cc}\phi$ chiral Lagrangians~\cite{Qiu:2020omj,Liu:2023lsg}. The input LEC values in these works are estimated from the $D\phi$ LECs, determined by fitting to lattice QCD of $D\phi$ interactions in, e.g., refs.~\cite{Liu:2012zya,Yao:2015qia}, via the heavy diquark-antiquark (HDA) symmetry~\cite{Savage:1990di}, where $D$ stands for ground-state pseudoscalar charmed mesons. Therefore, to obtain precise scattering information on the DCB spectroscopy, it is timely and necessary to carry out a lattice QCD calculation of the $B_{cc}\phi$ interactions.

In this work, we perform the first lattice QCD calculation of the $S$-wave scattering between the ground state $J^P=(1/2)^+$ DCBs and the Goldstone bosons. We use four $N_f=2+1$ gauge ensembles generated by the CLQCD collaboration~\cite{CLQCD:2023sdb}, with a lattice spacing $a=0.07746$ fm and two pion masses $M_{\pi} \sim 210$ and $300$ MeV. We focus on four single-channel elastic scattering processes: $\Omega_{cc}\bar{K}^{(-2,1/2)}$, $\Xi_{cc}K^{(1,1)}$, $\Xi_{cc}K^{(1,0)}$,  and $\Xi_{cc}\pi^{(0,3/2)}$, which are classified by the superscripts $(S,I)$ with $S$ and $I$ denoting strangeness and isospin, respectively. Pertinent interpolating operators for the DCBs and the Goldstone bosons are constructed, based on which the two-hadron operators that transform under the irreducible representation (irrep) $G_1^-$ of the discrete group $O_h^{(2)}$ are deduced, following the procedure detailed in ref.~\cite{Prelovsek:2016iyo}. For each scattering channel, finite-volume energy levels are extracted from the correlation functions of the corresponding operators. Our results reveal that the $\Xi_{cc}K^{(1,0)}$ channel is attractive at both pion masses, exhibiting downward energy shifts from the free thresholds, while the other three channels are repulsive. The scattering phase shifts are then obtained from the finite-volume energy levels via L\"uscher's finite volume method~\cite{Luscher:1990ux,Luscher:1991cf}. Consequently, we extract the $S$-wave scattering lengths and effective ranges, which are defined in the effective range expansion (ERE) parametrization of the threshold amplitude for the four channels. The scattering lengths are extrapolated to the physical point and are found to be consistent with the BChPT predictions~\cite{Meng:2018zbl,Liang:2023scp}, justifying the use of HDA symmetry in the estimate of the $B_{cc}\phi$ LEC values~\cite{Yan:2018zdt,Meng:2018zbl,Liang:2023scp}. In addition, a virtual state is found in the $\Xi_{cc}K^{(1,0)}$ channel through the pole analysis of the scattering amplitudes.

This manuscript is structured as follows. The lattice setup is briefly introduced in section~\ref{lattice.setup}. The finite-volume spectra are presented in~\ref{sec.spec}. Scattering analysis using L\"uscher formula is done in section~\ref{sec.scattering.ana}, where scattering lengths, effective ranges, and poles are discussed in detail. Our summary and outlook are presented in section~\ref{sec.sum}. Numerical results of the correlation functions are relegated to appendix~\ref{sec.res.cf}. Interpolators and masses needed for the estimate of coupled-channel effects from the singly charmed baryons $B_c$ and charm $D$ mesons are shown in appendix~\ref{appendix.otherchannels}. Appendix~\ref{sec.con} presents the quark contraction diagrams required for the computation of correlation functions of two-hadron systems.

\section{Lattice setup}\label{lattice.setup}
 
The results presented in this paper are based on the gauge configurations generated by the CLQCD Collaboration with $2+1$ dynamical quark flavors using the tadpole improved tree-level Symanzik gauge action and Clover fermion action~\cite{CLQCD:2023sdb}. We use four ensembles with the same lattice spacing $a=0.07746$~fm and two different pion masses $M_{\pi}\simeq 305$~MeV and $M_{\pi}\simeq207$~MeV. Details of the ensembles are listed in Table~\ref{tab:ensembles}. Among these ensembles, F32P21/F48P21 and F32P30/F48P30 are two couples that share the same pion mass but have different volumes to obtain more kinematic points in the finite-volume spectra, thereby rendering a more robust determination of the scattering parameters. The action of the valence charm quark is the Fermilab action~\cite{El-Khadra:1996wdx}, which controls discretization errors of $\mathcal{O}(am_c)^n$. The tuning of the parameters in the Fermilab action follows the method applied in Ref.~\cite{Liu:2009jc}. Specifically, the charm quark mass parameter $m_c$ and the anisotropy parameter $\nu$ in the action are tuned by requiring the mass of  $J/\psi$ to reproduce the physical value and the dispersion relation to match the continuum form.

\begin{table}[htb]
\begin{center}
\renewcommand{\tabcolsep}{0.4pc}
\begin{tabular}{c|c|c|c|c|c|c|c|c}
\hline
ID&$\beta$&$a$(fm)&$am_l$&$am_s$&$M_\pi$(MeV)&$L^3\times T$&$M_\pi L$&$N_{\rm cfgs.}$ \\
\hline
F32P30  &$6.41$  &$0.07746(18)$  &$-0.2295$ &$-0.2050$ &$303.9(0.6)$ &$32^3 \times 96$   &$3.818(8)$&$750$  \\
F48P30& $6.41$  &$0.07746(18)$  &$-0.2295$ &$-0.2050$ &$304.9(0.4)$   &$48^3 \times 96$ &$5.746(8)$&$359$  \\
F32P21&$6.41$   &$0.07746(18)$  &$-0.2320$   &$-0.2050$  &$208.1(1.9)$  &$32^3 \times 64$ &$2.614(24)$&$459$    \\
F48P21&$6.41$  &$0.07746(18)$ &$-0.2320$ &$-0.2050$ &$207.4(0.7)$  &$48^3 \times 96$  &$3.908(13)$&$222$   \\
\hline
\end{tabular}
\caption{Details of the four ensembles used in this work, including ensemble identifier (ID), gauge coupling $\beta$, lattice spacing $a$, dimensionless bare quark mass parameters for light and strange quarks ($am_l$, $am_s$), pion mass $M_\pi$, lattice volume $L^3 \times T$, dimensionless spatial size $M_\pi L$, and number of gauge configurations $N_{\rm cfgs.}$.}
\label{tab:ensembles}
\end{center}
\end{table}

Quark propagators are computed using the distillation quark smearing method~\cite{HadronSpectrum:2009krc}. The smearing operator is composed of a small number ($N_{\rm ev}$) of the eigenvectors associated with the $N_{\rm ev}$ lowest eigenvalues of the three-dimensional Laplacian on the HYP-smeared gauge field. The number of eigenvectors $N_{\rm ev}$ is $100$ for all four ensembles. We have verified that increasing $N_{\rm ev}$ to $200$ for the large-volume ensembles (F48P30 and F48P21) yields consistent energy levels. The statistical uncertainty is estimated using the bootstrap method with $4000$ samples. We have verified that this sample size is sufficient for the convergence of both the mean values and their statistical uncertainties.

\section{Finite-volume spectra\label{sec.spec}}
In this section, we present the spectra of the relevant single hadrons and two-hadron systems, which will be used to extract the scattering information via L\"uscher's method. We begin by detailing the construction of interpolating operators for both single- and two-hadron states, followed by a presentation of their corresponding numerical energy levels.

\subsection{Interpolating operators}
The interpolating operators for the Goldstone bosons ($\pi^+$, $K^+$ and $K^0$) and the ground-state $(1/2)^+$ DCBs ($\Xi_{cc}^{++,+}$, $\Omega_{cc}^+$) employed in this work are standard, which can be found in, e.g., refs.~\cite{Gattringer:2010zz,Liu:2009jc}. For easy reference and completeness, we provide their explicit forms in Appendix~\ref{appendix.otherchannels}.  In what follows, we focus on the construction of the two-hadron operators specific to our simulation.

For a hadron state of definite spatial momentum $\mathbf{p}$, the corresponding interpolator can be generically expressed as  
\begin{align}
\mathcal{O}_h(\mathbf{p},t)=\sum_{\mathbf{x}}\mathcal{O}_h(\mathbf{x},t)e^{i\mathbf{p}\cdot\mathbf{x}}\ ,\quad h\in\{\pi^\pm,K^\pm,K^0,\bar{K}^0,\Xi_{cc}^{++},\Xi_{cc}^{+},\Omega_{cc}^+\}\ ,\label{eq.one.particle}
\end{align}
with $\mathcal{O}_h(\mathbf{x},t)\equiv\mathcal{O}_h(x)$ given by eq.~\eqref{eq.interpolator.meson} for mesons and eq.~\eqref{eq.interpolator.baryon} for baryons. On the lattice, the momentum takes values of $\mathbf{p}=(n_x,n_y,n_z)\times (2\pi/L)$ with $n_{x,y,z}$ being integers. It is well acknowledged that the two-hadron states in the continuum can be classified by the irreducible representations of the SU(2) group, which are denoted by the total angular momentum $J=0,1/2,1,\cdots$. However, the continuum SU(2) group is reduced to the discrete double covering group $O^{(2)}$ of the cubic group $O$ on the lattice. When parity is taken into account, the $O^{(2)}$ is further promoted to $O^{(2)}_h=O^{(2)}\otimes Z_2$, with $Z_2=\{E={I}^2,{I}\}$ incorporating the spatial inversion ${I}$. Consequently, the construction of two-hadron operators from the one-hadron interpolators in eq.~\eqref{eq.one.particle} should be carried out according to the irrep $\Gamma=\{A_1^{\pm},A_2^{\pm},E^\pm, T_1^\pm, T_2^\pm, G_1^\pm, G_2^\pm, H^\pm\}$ of the discrete group $O^{(2)}_h$. A variety of approaches, such as the partial-wave~\cite{Berkowitz:2015eaa,Wallace:2015pxa,Prelovsek:2016iyo,Yan:2025jlq}, projection~\cite{Gockeler:2012yj,Prelovsek:2016iyo,Yan:2025jlq}, and helicity methods~\cite{Dudek:2012gj,Prelovsek:2016iyo}, have been proposed for constructing lattice operators that describe the scattering of particles with arbitrary spin.

Within the projection method, the two-hadron operator that transforms under the irrep $\Gamma$ of $O_h^{(2)}$ is defined by
\begin{align}
    \mathcal{O}_{p,\Gamma,r}^{h_1h_2} = \sum_{g\in O_h^{(2)}} T^{\Gamma}_{rr}(g)\, g \big[\mathcal{O}_{h_1}(\mathbf{p}_1)\mathcal{O}_{h_2}(\mathbf{p}_2) \big]g^{-1}\ ,\label{eq.two.particle}
\end{align}
where $T^\Gamma(g)$ stands for the matrix of the irrep $\Gamma$.
The summation runs over the $96$ elements in the $O_h^{(2)}=\{\cdots,g,\cdots\}$ group. The momenta $\mathbf{p}_1$ and $\mathbf{p}_2$ satisfy $\mathbf{p}_1+\mathbf{p}_2=0$ and $|\mathbf{p}_1|=|\mathbf{p}_2|\equiv p$, ensuring that the parity of the two-particle system is a good quantum number. The projection method leads to lattice operators that transform according to a given irrep $\Gamma$. In our case, the $S$-wave $B_{cc}\phi$ system with $J^P=(1/2)^-$ in the continuum corresponds to the irrep $G_1^-$ on the lattice. Due to the reduced symmetry of the lattice, the irrep $G_1^-$ contains mixing from higher angular momenta $J \ge 7/2$, which are sufficiently high to be ignored in our calculations. 

Eventually, for a given relative momentum modulus $p$, the scattering operators with definite $(S,I)$ quantum numbers are constructed as 
\begin{align}
\mathcal{O}_{p}
^{\Xi_{c c} K^{(1,0)}}
&= \frac{1}{\sqrt{2}}\sum_{\alpha, \mathbf{p}_{\mathbf{1}}, \mathbf{p}_{\mathbf{2}}} C_{\alpha, \mathbf{p}_{\mathbf{1}}, \mathbf{p}_{\mathbf{2}}}
\left(
 \mathcal{O}_{\Xi_{c c}^{++}, \alpha}(\mathbf{p}_{\mathbf{1}}) \mathcal{O}_{K^0}(\mathbf{p}_{\mathbf{2}})- \mathcal{O}_{\Xi_{c c}^{+}, \alpha}(\mathbf{p}_{\mathbf{1}}) \mathcal{O}_{K^{+}}(\mathbf{p}_{\mathbf{2}})
\right)\ ,\notag\\
\mathcal{O}_{p}
^{\Xi_{c c} K^{(1,1)}}
&=\sum_{\alpha, \mathbf{p}_{\mathbf{1}}, \mathbf{p}_{\mathbf{2}}} C_{\alpha, \mathbf{p}_{\mathbf{1}}, \mathbf{p}_{\mathbf{2}}}
\left(
\mathcal{O}_{\Xi_{c c}^{++}, \alpha}(\mathbf{p}_{\mathbf{1}}) \mathcal{O}_{K^+}(\mathbf{p}_{\mathbf{2}})
\right)\ ,\notag\\
\mathcal{O}_{p}
^{\Omega_{c c} \bar{K}^{(-2,1/2)}}
&= \sum_{\alpha, \mathbf{p}_{\mathbf{1}}, \mathbf{p}_{\mathbf{2}}} C_{\alpha, \mathbf{p}_{\mathbf{1}}, \mathbf{p}_{\mathbf{2}}}
\left(
\mathcal{O}_{\Omega_{c c}^{+}, \alpha}(\mathbf{p}_{\mathbf{1}}) \mathcal{O}_{\bar{K}^0}(\mathbf{p}_{\mathbf{2}})
\right),
\notag\\
\mathcal{O}_{p}
^{\Xi_{c c} \pi^{(0,3/2)}}
&=\sum_{\alpha, \mathbf{p}_{\mathbf{1}}, \mathbf{p}_{\mathbf{2}}} C_{\alpha, \mathbf{p}_{\mathbf{1}}, \mathbf{p}_{\mathbf{2}}}
\left(
\mathcal{O}_{\Xi_{c c}^{++}, \alpha}(\mathbf{p}_{\mathbf{1}}) \mathcal{O}_{\pi^+}(\mathbf{p}_{\mathbf{2}})
\right)\ .\label{eq.our.twoparticle.ops}
\end{align}
Here, $\alpha$ is the Dirac index of the doubly charmed baryon field. The coefficients $C_{\alpha, \mathbf{p}_{\mathbf{1}}, \mathbf{p}_{\mathbf{2}}}$ are chosen to ensure the operators transform under the irrep $G_1^-$. In Table~\ref{tab:two_op_coefs} of Appendix~\ref{appendix.otherchannels} , we list the values of $C_{\alpha, \mathbf{p}_{\mathbf{1}}, \mathbf{p}_{\mathbf{2}}}$ for the lowest three momenta $p^2=0,1,2$ (in units of $(2\pi/L)^2$).

\subsection{Spectra of the single particles}

The masses of the single particles are obtained from the correlation functions of the respective interpolating operators defined in \cref{eq.one.particle}:  
\begin{align}
C_h(\mathbf{p},t) = \sum_{t_{\rm src}
}\langle \mathcal{O}_h(\mathbf{p},t + t_{\rm src} )\mathcal{O}_h^\dagger(\mathbf{p},t_{\rm src})\rangle \ .
\end{align}
The source time $t_{\rm src}$ is summed over all time slices to enhance the statistics. The dispersion relation $E^2 = m_0^2 + c^2\mathbf{p}^2$  is investigated by calculating the single-particle energies at the five lowest momenta on lattice: $\mathbf{p}=$ $(0,0,0)$, $(0,0,1)$, $(0,1,1)$, $(1,1,1)$, $(0,0,2)$ in units of $\frac{2\pi}{L}$. The effective masses for pion, kaon, $\Xi_{cc}$ and $\Omega_{cc}$ at the five momenta are displayed in~\cref{fig:pion.meff,fig:kaon.meff,fig:Xicc.meff,fig:Omegacc.meff} of \cref{sec.res.cf}, respectively. We fit the correlation functions to an exponential form $C(t)=Ae^{-Et}$ for the baryons and a cosh function $C(t)=A\cosh[{E(T-\frac{t}{2})}]$ for the mesons to obtain the energies. The masses of the single particles at zero momentum are listed in table~\ref{tab:single.particle.meff}. The results of dispersion relation fits are given in \cref{tab:DisRel.FitRslts}. The fitted parameter $c^2$ deviates slightly from unity due to lattice artifacts, but its proximity to one indicates that these effects are small. The rest masses $m_0$ extracted from the dispersion relation fits are in perfect agreement with those from direct correlator fits (c.f. \cref{tab:single.particle.meff}). 

\begin{table}[htb]
\begin{center}
\begin{tabular}{c|c|c|c|c}
\hline
      &     $\pi$&       $K$& $\Xi_{cc}$& $\Omega_{cc}$\\
\hline    
F32P30&  $303.9(6)$ & $523.0(4)$ & $3632.4(9)~~$& $3713.7(6)~~$
\\
\hline 
F48P30&  $304.9(4)$ &$524.1(3)$ &$3637.2(1.3)$ & $3717.9(1.0)$
\\
\hline 
F32P21& $208.1(1.9)$ &$491.7(7)$ &$3605.6(1.4)$ & $3693.8(8)~~$
\\
\hline 
F48P21&  $207.4(7)$ &$491.1(3)$ &$3607.6(1.8)$ & $3699.7(1.0)$
\\
\hline
\end{tabular}
\caption{Masses of single particles, in units of MeV.}
\label{tab:single.particle.meff}
\end{center}
\end{table}

\begin{table}[ht] 
\begin{center}
\begin{tabular}{c |c|c|c|c}
\hline
Particle & Ensemble & $m_0$ (MeV) & $c^2$ & $\chi^2/{\rm dof}$ \\
\hline
\multirow{4}{*}{$\pi$} & F32P30 & $303.96(60)$ & $1.0069(42)$ & $0.69$ \\
 & F48P30 & $304.87(39)$ & $1.0075(22)$ & $1.02$ \\
 & F32P21 & $208.50(1.90)$ & {$1.0582(94)$} & $0.73$ \\
 & F48P21 & $207.74(69)$ & {$1.0157(37)$} & {$1.58$} \\
\hline
\multirow{4}{*}{$K$} & F32P30 & $523.00(43)$ & $1.0026(23)$ & {$1.47$} \\
 & F48P30 & $524.07(28)$ & $1.0034(15)$ & $0.53$ \\
 & F32P21 & $491.73(69)$ & $1.0097(52)$ & $0.26$ \\
 & F48P21 & $491.09(33)$ & $1.0081(25)$ & $0.63$ \\
\hline
\multirow{4}{*}{$\Xi_{cc}$} & F32P30 & $3633.0(8)~~$ & $0.9915(61)$ & {$1.70$} \\
 & F48P30 & $3636.9(1.2)$ & $0.9640(190)$ & $0.38$ \\
 & F32P21 & $3605.5(1.4)$ & $1.0060(100)$ & $0.48$ \\
 & F48P21 & $3608.0(1.6)$ & $1.0520(220)$ & $0.56$ \\
\hline
\multirow{4}{*}{$\Omega_{cc}$} & F32P30 & $3713.9(6)~~$ & $0.9747(37)$ & $0.92$ \\
 & F48P30 & $3717.9(1.0)$ & $0.9670(200)$ & $0.44$ \\
 & F32P21 & $3693.8(8)~~$ & $0.9801(51)$ & $0.11$ \\
 & F48P21 & $3699.5(1.0)$ & $0.9980(130)$ & $1.16$ \\
\hline
\end{tabular}
\caption{Results of single-particle masses $m_0$ and dispersion-relation coefficients $c^2$, obtained by fitting the single particle energies at the five lowest momenta to the dispersion relation $E^2=m_0^2+c^2\mathbf{p}^2$. }
\label{tab:DisRel.FitRslts}
\end{center}
\end{table}

\subsection{Finite-volume spectra of the two-hadron systems\label{sec.4pt.cf.el}}

The energies of the scattering systems can be extracted from the correlation functions of the two-hadron operators. For each channel under consideration, a correlation matrix is constructed as 
\begin{align}
\mathbf{C}_{i j}(t)=\sum_{t_{\rm s r c}}\left\langle\mathcal{O}_i\left(t+t_{\rm s r c}\right) \mathcal{O}_j^{\dagger}\left(t_{\rm s r c}\right)\right\rangle \ ,
\end{align}
where $\mathcal{O}_i$ and $\mathcal{O}_j$ are given in eq.~\eqref{eq.our.twoparticle.ops}, and the subscripts $i,j$ label different values of momenta $p$. For the set of momenta $p=\{0, 1, \sqrt{2}\}(2\pi/L)$, the correlation function is a $3\times 3$ matrix. The quark contraction diagrams for the four channels, $\Omega_{cc}\bar{K}^{(-2,1/2)}$, $\Xi_{cc}K^{(1,1)}$, $\Xi_{cc}K^{(1,0)}$ and $\Xi_{cc}\pi^{(0,3/2)}$, required in our simulation, are shown in \cref{fig:OmgccKbar_m2_0dot5_contraction,fig:XiccK_1_1_contraction,fig:XiccK_1_0_contraction,fig:XiccPi_contraction} of~\cref{sec.con}, respectively.

The finite-volume energy spectra of two-hadron systems can be determined through a systematic analysis of the correlation function matrices $\mathbf{C}$ using the variational technique suggested in ref.~\cite{Luscher:1990ck}. Specifically, the generalized eigenvalue problem (GEVP)~\cite{Michael:1985ne,Luscher:1990ck,Blossier:2009kd} is solved with a fixed time slice $t_0$. In this study, we choose $t_0=3$ in lattice units and have verified that the resulting energies are stable against variations of this choice:
\begin{align}
    \mathbf{C}(t) v^{(n)}(t)=\lambda_n(t) \mathbf{C}\left(t_0\right) v^{(n)}(t)\ ,\quad n=1,2,3\ ,
\end{align}
where $\lambda_{n}$ and $v^{(n)}$ denote eigenvalues and eigenvectors, respectively. The energies $E_n$ are extracted from the eigenvalues $\lambda_n(t)$ by 
\begin{align}
    \lambda_n(t)=(1-\left.A_n\right) e^{-E_n\left(t-t_0\right)}+A_n e^{-E_n^{\prime}\left(t-t_0\right)}\ ,\label{eq.lambda.two.exp}
\end{align}
where $A_n$, $E_n$, and $E_n^{\prime}$ are free parameters. $E_n^{\prime}$ accounts for high-lying energy levels not covered by GEVP analysis. Note that $\lambda_n\to 1$ as $t\to t_0$, and the first term dominates for large $t$ since $E_n^\prime >E_n$. We extract the energy levels by performing two-exponential fits to the temporal dependence of the eigenvalues $\lambda_n(t)$ using \cref{eq.lambda.two.exp}. To visualize the fit quality and validate our choice of fit ranges, we present plots of the rescaled eigenvalues for the four two-hadron systems in~\cref{fig:OmegaccKbar.meff,fig:XiccK11.meff,fig:XiccK10.meff,fig:XiccPi.meff} of~\cref{sec.res.cf}.

\begin{figure}[htbp]
\centering
\begin{subfigure}[b]{0.85\textwidth}
\includegraphics[width=\textwidth]{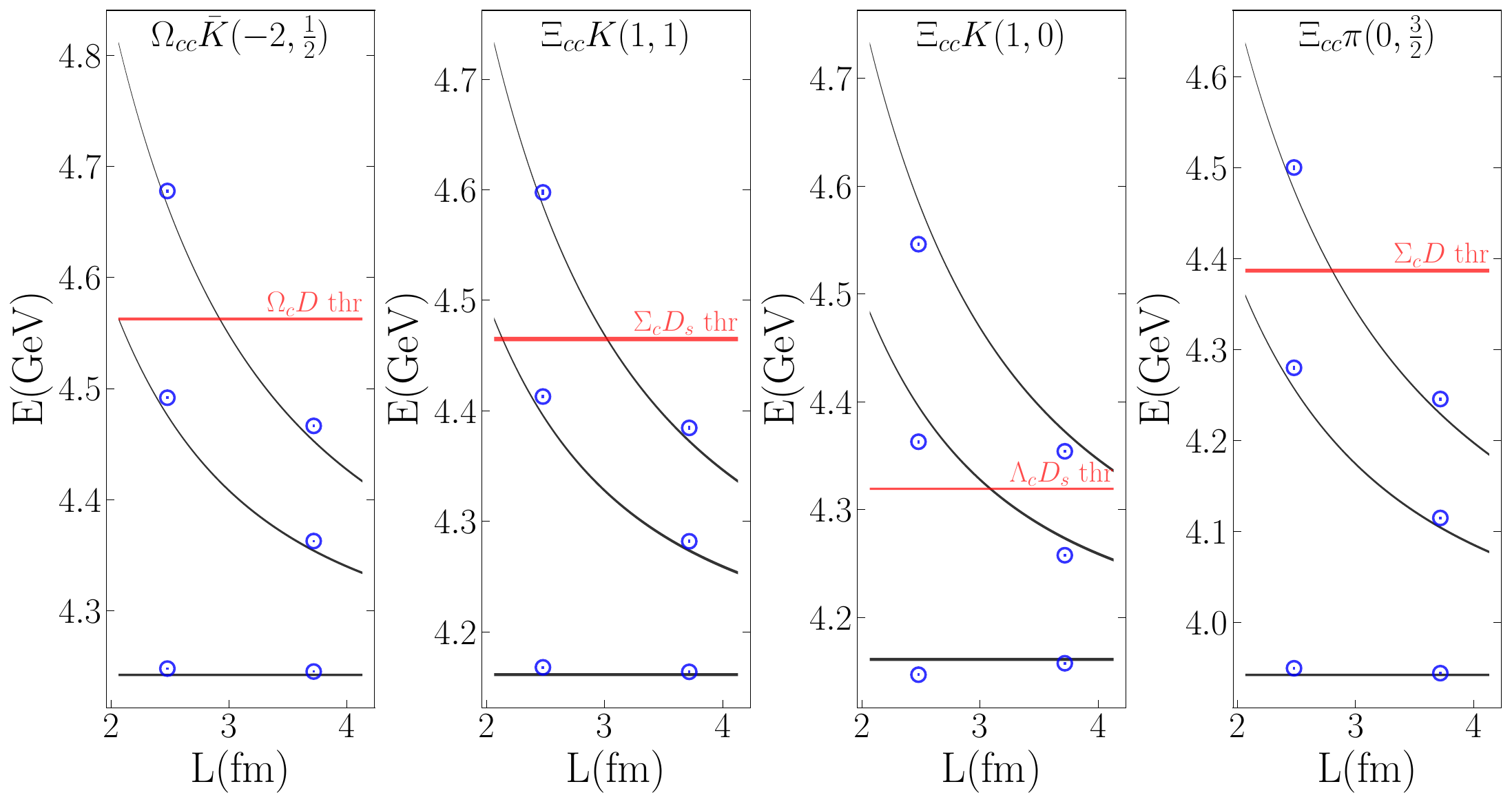}
\caption{}
\end{subfigure}
\begin{subfigure}[b]{0.85\textwidth}
\includegraphics[width=\textwidth]{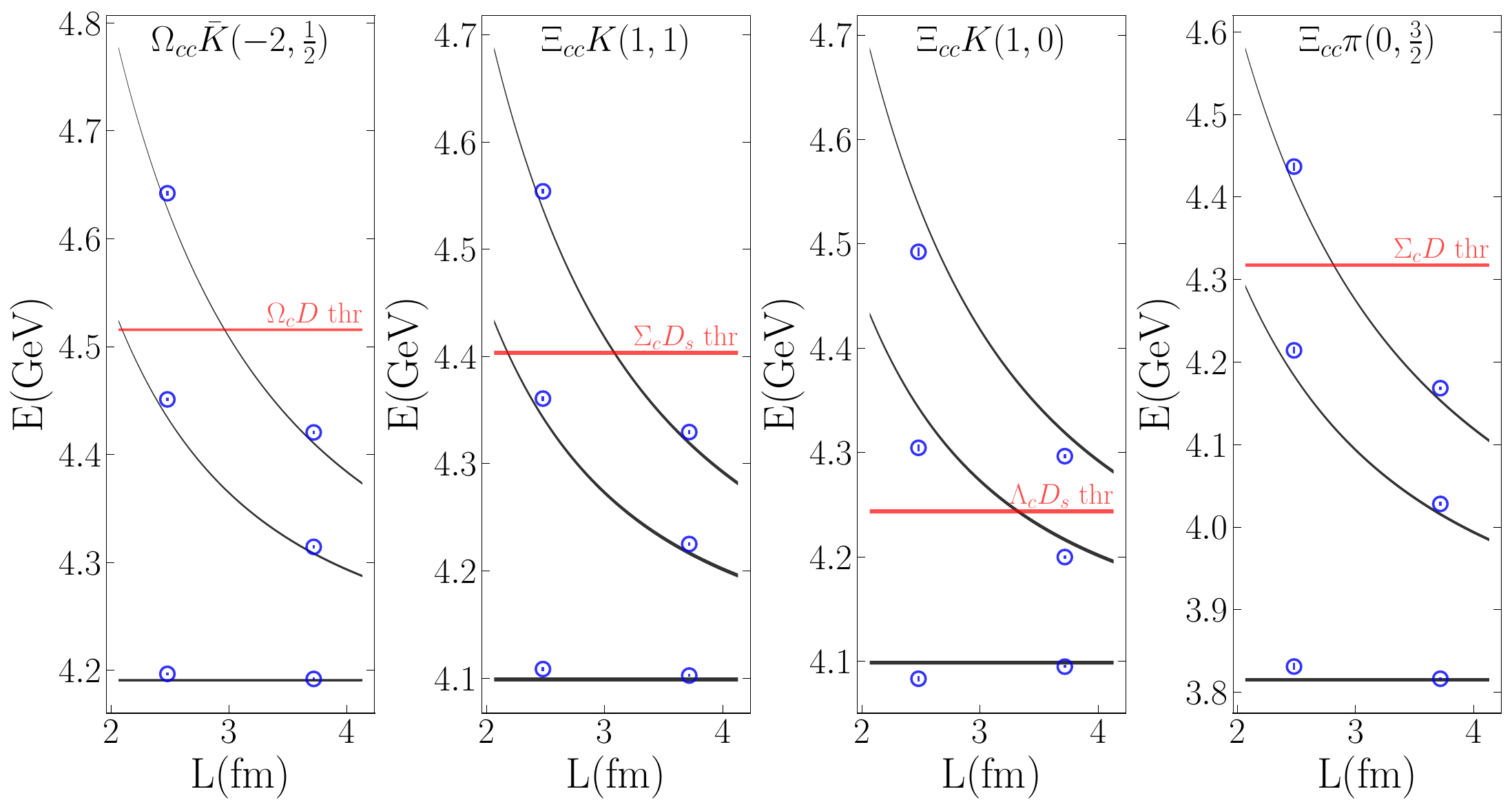}
\caption{}
\end{subfigure}
\caption{Energy levels of two-hadron systems for the four single channels. (a) $M_\pi \sim 300$ MeV; (b) $M_\pi \sim 210$ MeV. The blue data points are the finite-volume energies, and the black bands indicate free energies of the non-interacting threshold with different momenta.}
\label{fig:energy.levels}
\end{figure}

Figure~\ref{fig:energy.levels} shows the obtained energy levels represented by blue points alongside the non-interacting thresholds depicted as black bands for comparison. The $\Xi_{cc}K^{(1,0)}$ channel displays consistent negative energy shifts across both pion mass ensembles; namely, finite-volume energy levels reside below the non-interacting thresholds. This implies that the $\Xi_{cc}K^{(1,0)}$ system is attractive and may form a bound state or virtual state in the infinite-volume limit, necessitating the scattering analysis detailed in section~\ref{sec.scattering.ana}. In contrast, the energy levels of the other three channels are slightly above the respective thresholds, which is indicative of repulsive interactions.

Before ending this section, it should be noted that the above single-channel description is strictly valid only for the systems composed of $B_{cc}$ ($ccq$) and Goldstone bosons $\phi$ ($q\bar{q}$). In fact, the two-hadron systems $B_{cc}\phi$ can also couple to channels comprising a singly charmed baryon $B_c$ ($cqq$) and a charmed meson $D$ ($c\bar{q}$), since the total quark content $ccqq\bar{q}$ remains conserved. For instance, the $\Omega_{cc}\bar{K}^{(-2,\frac{1}{2})}$ mixes with $\Omega_c D^{(-2,\frac{1}{2})}$, etc. In~\cref{fig:energy.levels}, we plot the relevant $B_c D$ thresholds to transparently assess the influence of the $B_c D$ channels on our energy levels obtained under the single-channel approximation. The required interpolators for $B_c\in\{\Sigma_{c}^{++},\Lambda_c^+,\Omega_{c}^0\}$ and $D\in\{D^0,D_s^+\}$ are constructed and shown in~\cref{eq.Bc.D.interpolators}. Their masses at unphysical pion masses, $\sim 210$ and $300$~MeV, are estimated using gauge ensembles F48P30 and F48P21, with the results compiled in~\cref{tab:meff.others} in~\cref{appendix.otherchannels}. The energy levels above the $B_c D$ thresholds are excluded from our subsequent scattering analysis in section~\ref{sec.scattering.ana}, where the contributions of $B_c D$ channels cannot be ignored.

\section{Scattering analysis\label{sec.scattering.ana}} 

\subsection{L\"uscher's formula and effective range expansion}

The discrete energy levels of the interacting $B_{cc}\phi$ system, as established in the preceding section, can be expressed as
\begin{align}
E_p=\sqrt{m_{B_{cc}}^2+p^2}+\sqrt{m_{\phi}^2+p^2}\ ,\label{eq.EL.p}
\end{align}
where $p$ is no longer constrained to the original relative momentum mode $n\frac{2\pi}{L}$ ($n=|\mathbf{n}|$ and $\mathbf{n}\in \mathbb{Z}^3$) in the non-interacting case. The shift of $p$ from its free mode encodes the underlying scattering information of the finite-volume $B_{cc}\phi$ spectrum on the lattice, which can be translated to infinite-volume matrix elements by means of the L\"uscher's formula~\cite{Luscher:1990ux,Luscher:1991cf} and its generalizations~\cite{Briceno:2014oea}. 

The L\"uscher quantization condition relates the finite-volume energy spectra, modified by interactions, to the infinite-volume scattering amplitudes. In our single-channel case, the $S$-wave scattering amplitude can be parameterized as
\begin{align}
T=\frac{1}{p\cot\delta_0-ip}\ ,\label{eq:scat.amp}
\end{align}
where $\delta_0$ denotes the $S$-wave phase shift. Note that we employ a normalization such that the $S$-matrix element is associated with the $T$ matrix by $S=1+2i p T$ and the unitarity relation reads ${\rm Im}\, T= p \,|T|^2$. Furthermore, with this normalization, $T$ is identical to the amplitude $f_{0^+}$ specified by eq.~(2.26) of ref.~\cite{Liang:2023scp}, i.e., $T=f_{0^+}$. Consequently, the L\"uscher quantization condition has the well-known form
\begin{equation}
p \cot \delta_0(p)=\frac{2}{L \sqrt{\pi}} \mathcal{Z}_{00}\left(1 ; q^2\right)\ ,\quad q=\frac{pL}{2\pi}\ ,
\label{eq:Luscher.formula.swave}
\end{equation}
where $\mathcal{Z}_{00}(s;q^2)$ is the L\"uscher Zeta function defined in~ref.~\cite{Luscher:1990ux}. Since $p$ is obtainable from the interacting energy levels via eq.~\eqref{eq.EL.p}, one may readily obtain the phase shift $\delta_0(p)$ using eq.~\eqref{eq:Luscher.formula.swave}.

At low energies, the scattering amplitude~\eqref{eq:scat.amp} can be expanded as a series of $p^2$, accompanied by threshold parameters, which is referred to as the ERE. Traditionally, the ERE of the inverse of the scattering amplitude at threshold reads
\begin{equation}
{\rm Re}\left(\frac{1}{T}\right)=p \cot \delta_0=\frac{1}{a_{0}}+\frac{1}{2} r_{0} p^2+\sum_{n=2}^\infty v_n p^{2n}\ ,
\label{eq:ere}
\end{equation}
where the threshold parameters $a_{0}$, $r_{0}$ and $v_n$ ($n\geq 2$) are referred to as the scattering length, effective range and shape parameters, in order. In the next subsection, we are going to determine the scattering lengths of the four single channels with the help of~\cref{eq:Luscher.formula.swave,eq:ere}.

\subsection{Scattering lengths}

\begin{figure}[htbp] 
\begin{subfigure}[b]{1\linewidth}
\centering
\includegraphics[width=1\linewidth]{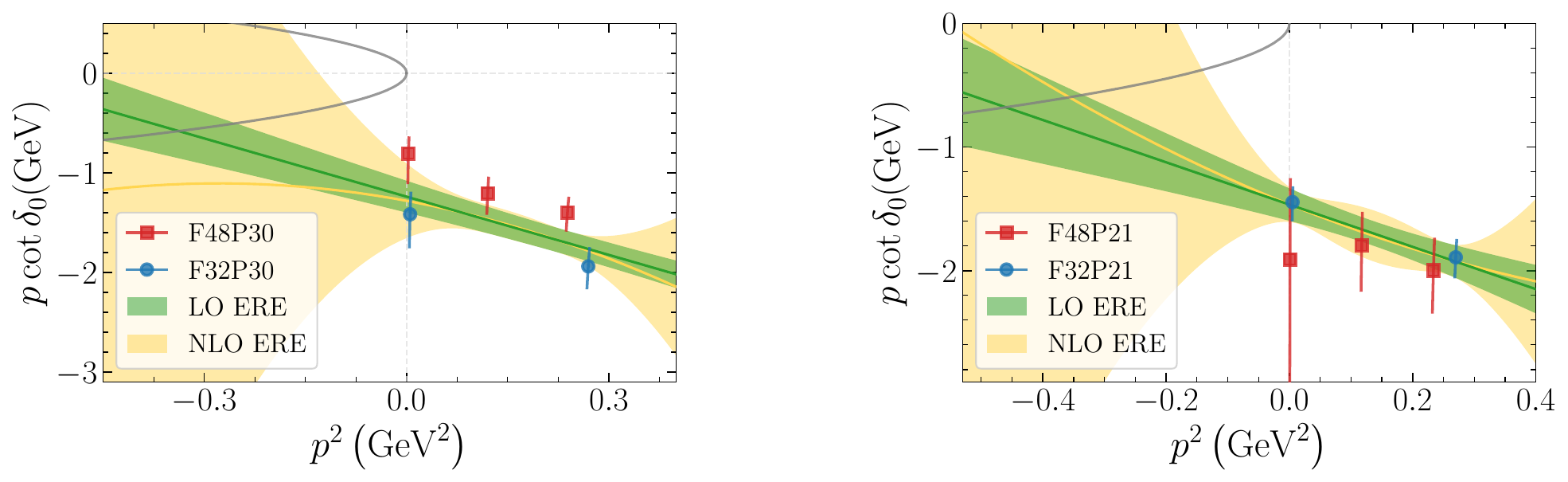} 
\caption{$\Omega_{cc}\bar{K}^{(-2,\frac{1}{2})}$} 
\label{fig:ERE.shifted.a} 
\end{subfigure}
\hfill
\begin{subfigure}[b]{1\linewidth}
\centering
\includegraphics[width=1\linewidth]{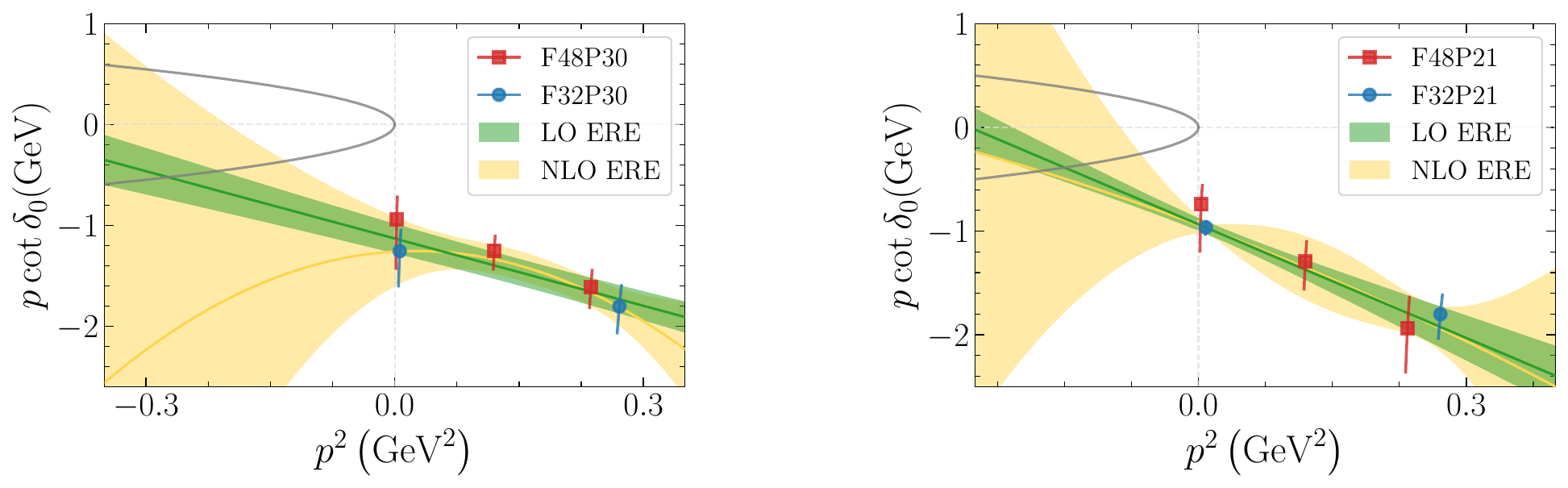} 
\caption{$\Xi_{cc}K^{(1,1)}$} 
\label{fig:ERE.shifted.b} 
\end{subfigure} 
\hfill
\begin{subfigure}[b]{1\linewidth}
\centering
\includegraphics[width=1\linewidth]{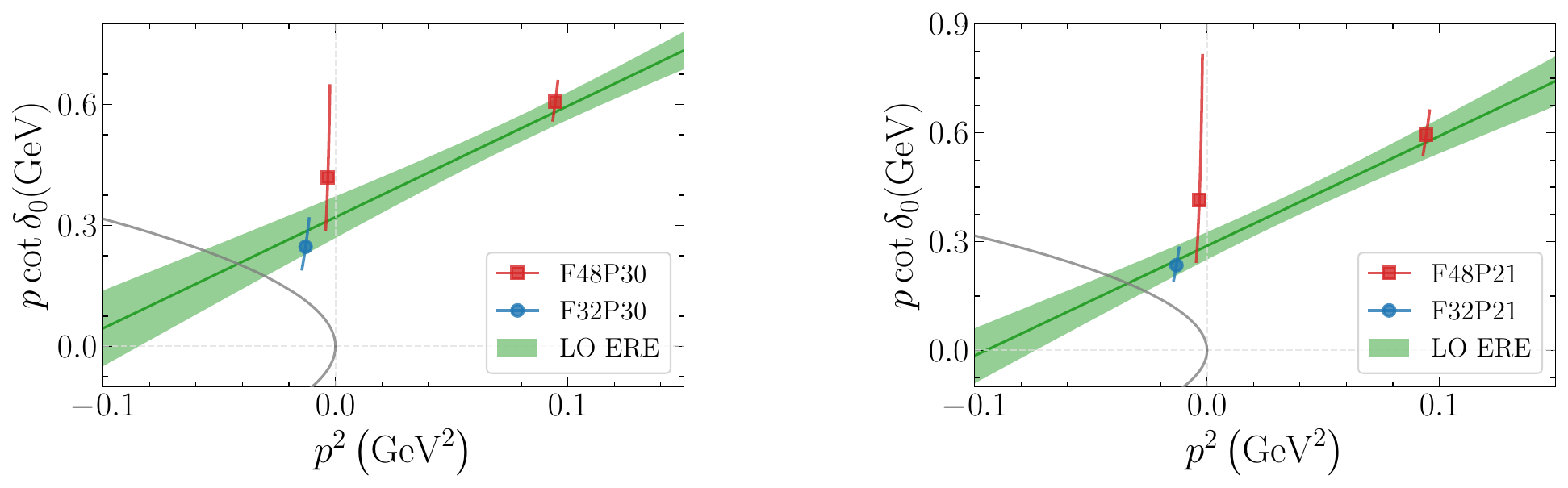} 
\caption{$\Xi_{cc}K^{(1,0)}$} 
\label{fig:ERE.shifted.c} 
\end{subfigure}
\hfill
\begin{subfigure}[b]{1\linewidth}
\centering
\includegraphics[width=1\linewidth]{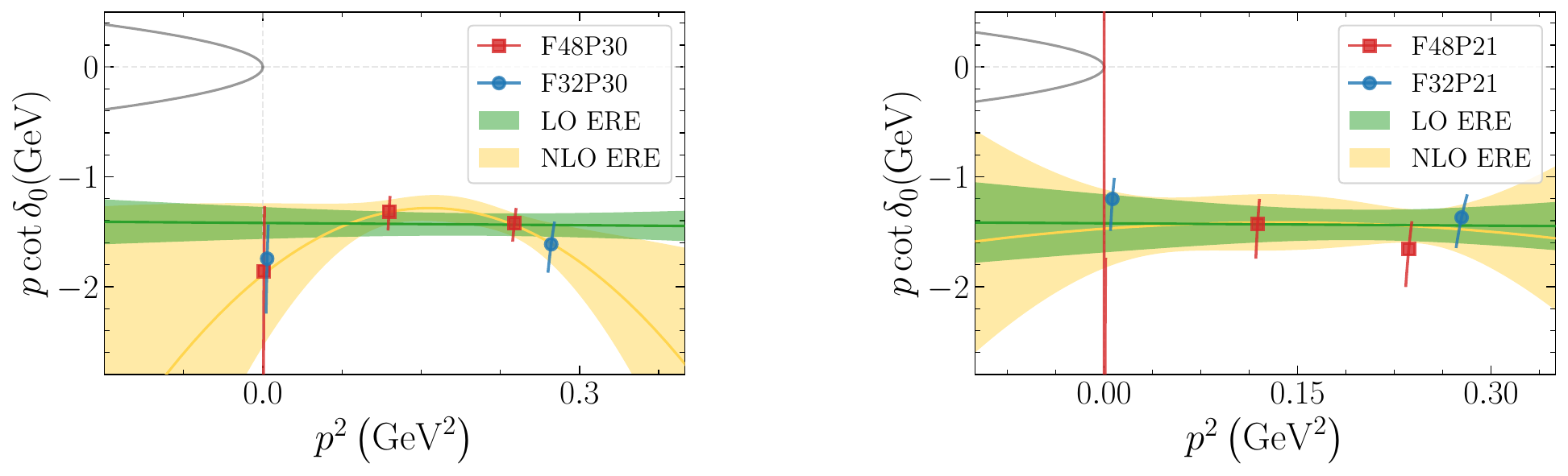} 
\caption{$\Xi_{cc}\pi^{(0,\frac{3}{2})}$} 
\label{fig:ERE.shifted.d} 
\end{subfigure} 
\caption{ERE fit results truncated at $\mathcal{O}(p^2)$ (green bands) and $\mathcal{O}(p^4)$ (yellow bands) for four single channels using DRC energy levels. The left and right panels correspond to pion masses $M_{\pi}\sim300$ MeV and 210 MeV, respectively. The red and blue points show $p\cot\delta_0$ extracted via L\"uscher's formula (eq.~(\ref{eq:Luscher.formula.swave})), while the grey solid curve is $ip=\pm |p|$ versus $p^2$.}
\label{fig:ERE.shifted}
\end{figure}

By virtue of the L\"uscher formula~\eqref{eq:Luscher.formula.swave}, the values of $p\cot\delta_0$ at various momenta $p$ can be determined from the interacting energy levels obtained in~\cref{sec.4pt.cf.el}. There are two points to be discussed before extracting $p\cot\delta_0$. The first one is that we shift the finite-volume energy levels according to 
\begin{equation}
E_p \to E_p+  \left[E_{\phi}^{\rm cont.}(p)+E_{B_{cc}}^{\rm cont.}(p)\right]-\left[E_{\phi}^{\rm latt.}(p)+E_{B_{cc}}^{\rm latt.}(p) \right]\ ,
\label{eq:dispersion.relation.shift}
\end{equation}
where $E^{\rm cont.}_{h}(p)$ ($h\in\{\phi,B_{cc}\}$) are the energies calculated from the continuum dispersion relation $E^{\rm cont.}_{h}(p)=\sqrt{m_h^2+p^2}$, while $E^{\rm latt.}_{h}(p)$ corresponds to the single-particle energies computed on the lattice. Such a shift is implemented to alleviate the lattice artifacts arising from discretization effects following~\cite{Padmanath:2022cvl,Prelovsek:2020eiw, Piemonte:2019cbi, Xing:2022ijm}. Specifically, since the L\"uscher formalism assumes continuum kinematics, the deviation of the lattice dispersion relation from the continuum form (where $c^2 \neq 1$) can introduce systematic errors. This correction aligns the kinematic behavior of the lattice data with the continuum expectation, thereby ensuring consistency with the quantization condition. The other point concerns the selection of the energy levels. That is, only the energy levels below the $B_cD$ thresholds are kept in our scattering analysis, as already explained at the end of~\cref{sec.4pt.cf.el}. Our final results of $p\cot\delta_0$, derived from energy levels with the dispersion relation correction (DRC) implemented via eq.~\eqref{eq:dispersion.relation.shift}, are shown as red squares and blue dots in figure~\ref{fig:ERE.shifted}. 

We are now in the position to extract the $S$-wave scattering parameters. To properly account for the statistical correlations between the energy levels, we perform a correlated fit by minimizing a $\chi^2$ function defined at the level of squared momenta,
\begin{equation}
\chi^2(a_0, r_0, \dots) = \sum_{i, j} \left[ p^2(E_i) - p_{\rm ERE}^2(i, a_0, r_0, \dots) \right] (C_{\rm cov}^{-1})_{ij} \left[ p^2(E_j) - p_{\rm ERE}^2(j, a_0, r_0, \dots) \right].
\label{eq:chi2_corr_def}
\end{equation}
Here, the indices $i, j$ run over all energy levels included in the fit for a given scattering channel and pion mass. The quantity $p^2(E_i)$ is the squared momentum determined from the $i$-th lattice energy level $E_i$ by numerically solving eq.~(\ref{eq.EL.p}), $p_{\rm ERE}^2(i,a_0, r_0, \dots )$ is the solution of the L\"uscher quantization condition~\eqref{eq:Luscher.formula.swave} with ERE parametrization ~\eqref{eq:ere}, and $C_{\rm cov}$ is the covariance matrix of the squared momenta.

By truncating the ERE at $\mathcal{O}(p^2)$ (denoted as leading order (LO) hereafter), we obtain the scattering length $a_0$ and effective range $r_0$. The resulting values of $a_0$ and $r_0$ are collected in table~\ref{tab:ERE.FitRslts.UpToP2}. The labels \lq\lq Raw" and \lq\lq DRC" in the table denote the $p\cot\delta_0$ datasets derived from energy levels without and with DRC, respectively. It can be found that the Raw and DRC-applied results are consistent with each other within 1-$\sigma$ uncertainties, indicating the impact of DRC on $a_0$ and $r_0$ is mild. In particular for $a_0$, it is predominantly determined by the lowest-lying two-particle energy levels nearest to the non-interacting threshold, which have zero momentum $p=\left|\mathbf{p}_{\mathbf{1}, \mathbf{2}}\right|=0$ with vanishing DRC term $c^2p^2$ in the dispersion relation. Nevertheless, we adopt the values from fits to the DRC-applied data as our final results for both threshold parameters. Based on the fitted values, our LO ERE predictions of $p\cot\delta_0$ are presented as solid green lines with bands in figure~\ref{fig:ERE.shifted}.

\begin{table}[ht] 
\begin{center}
\begin{tabular}{c c|c|c|c|c|c}
\hline
$(S,I)$ & Processes & $M_{\pi}$ (MeV) & Data type & $a_0$ (fm) & $r_0$ (fm) & $\chi^2/{\rm dof}$ \\
\hline
\multirow{4}{*}{$(-2,\frac{1}{2})$} &\multirow{4}{*}{$\Omega_{cc} \bar{K} \to \Omega_{cc} \bar{K}$} & \multirow{2}{*}{$300$} & Raw & $-0.172(13)$ & $-1.04(10)$ & $1.11$ \\
 & & & DRC & $-0.162(20)$ & $-0.77(17)$ & $1.15$ \\
\cline{3-7}
 & & \multirow{2}{*}{$210$} & Raw & $-0.130(11)$ & $-0.40(19)$ & $0.67$ \\
 & & & DRC & $-0.136(12)$ & $-0.67(25)$ & $0.15$ \\
\hline
\multirow{4}{*}{$(1,1)$} &\multirow{4}{*}{$\Xi_{cc} K \to \Xi_{cc} K$ } & \multirow{2}{*}{$300$} & Raw & $-0.184(16)$ & $-1.09(12)$ & $0.30$ \\
 & & & DRC & $-0.177(22)$ & $-0.88(16)$ & $0.48$ \\
\cline{3-7}
 & & \multirow{2}{*}{$210$} & Raw & $-0.211(14)$ & $-0.88(14)$ & $0.45$ \\
 & & & DRC & $-0.212(14)$ & $-1.44(29)$ & $0.53$ \\
\hline
\multirow{4}{*}{$(1,0)$} &\multirow{4}{*}{$\Xi_{cc} K \to \Xi_{cc} K$ } & \multirow{2}{*}{$300$} & Raw & $0.640(100)$ & $0.98(14)$ & $1.08$ \\
 & & & DRC & $0.630(100)$ & $1.09(19)$ & $1.05$ \\
\cline{3-7}
 & & \multirow{2}{*}{$210$} & Raw & $0.690(90)$ & $1.29(10)$ & $0.67$ \\
 & & & DRC & $0.697(90)$ & $1.19(20)$ & $0.69$ \\
\hline
\multirow{4}{*}{$(0,\frac{3}{2})$} &\multirow{4}{*}{$\Xi_{cc} \pi \to \Xi_{cc} \pi$ } & \multirow{2}{*}{$300$} & Raw & $-0.166(11)$ & $-0.23(11)$ & $4.29$ \\
 & & & DRC & $-0.140(14)$ & $-0.03(19)$ & $1.54$ \\
\cline{3-7}
 & & \multirow{2}{*}{$210$} & Raw & $-0.149(21)$ & $0.52(28)$ & $1.17$ \\
 & & & DRC & $-0.143(24)$ & $-0.03(45)$ & $1.02$ \\
\hline
\end{tabular}
\caption{Single-channel $S$-wave scattering parameters ($a_0$, $r_0$), and the corresponding $\chi^2/{\rm dof}$ obtained from LO ERE fits for four distinct scattering systems at pion masses $M_{\pi} \sim 300$ MeV and $210$ MeV. The raw (uncorrected) and DRC results are compared to quantify deviations in the relativistic dispersion relation of heavy baryons within L\"uscher's finite-volume framework.
}
\label{tab:ERE.FitRslts.UpToP2}
\end{center}
\end{table}

To assess the influence of the ERE truncation on the scattering lengths, we perform additional fits by employing the ERE expression up to $\mathcal{O}(p^4)$ (labeled as next-to-leading order (NLO) hereafter). Our NLO ERE predictions are shown as yellow lines with bands in figure~\ref{fig:ERE.shifted}. Due to limited data points and one additional free parameter $v_2$, the NLO fits lead to substantially larger uncertainties compared to the LO fits, as can be seen in the figure. In table~\ref{tab:ERE.FitRslts.P2vsP4}, we compare the results of the scattering length $a_0$ obtained from the LO and NLO fits. One finds that the scattering length $a_0$ remains stable  under different ERE truncation orders, with deviation within $1\sigma$ for most channels. This truncation insensitivity of $a_0$ aligns with its dominant contribution from near-threshold energy levels. We do not show the effective range $r_0$ and shape parameter $v_2$ in table~\ref{tab:ERE.FitRslts.P2vsP4} due to their large statistical errors. Note also that the dataset for the $\Xi_{cc}K^{(1,0)}$ channel is insufficient for us to conduct an NLO ERE fit with three free parameters.

\begin{table}[ht] 
\begin{center}
\begin{tabular}{c c|c|c|c|c|c}
\hline
$(S,I)$ & Processes & $M_{\pi}$ (MeV) & Data type & $a_0$ (LO) & $a_0$ (NLO) & Difference\\
\hline
\multirow{4}{*}{$(-2,\frac{1}{2})$} &\multirow{4}{*}{$\Omega_{cc} \bar{K} \to \Omega_{cc} \bar{K}$} & \multirow{2}{*}{$300$} & Raw & $-0.172(13)$ & $-0.164(23)$ & $<1\sigma$ \\
 & & & DRC & $-0.162(20)$ & $-0.159(25)$ & $<1\sigma$ \\
\cline{3-7}
 & & \multirow{2}{*}{$210$} & Raw & $-0.130(11)$ & $-0.135(13)$ & $<1\sigma$ \\
 & & & DRC & $-0.136(12)$ & $-0.136(13)$ & $<1\sigma$ \\
\hline
\multirow{4}{*}{$(1,1)$} &\multirow{4}{*}{$\Xi_{cc} K \to \Xi_{cc} K$ } & \multirow{2}{*}{$300$} & Raw & $-0.184(16)$ & $-0.170(28)$ & $<1\sigma$ \\
 & & & DRC & $-0.177(22)$ & $-0.163(29)$ & $<1\sigma$ \\
\cline{3-7}
 & & \multirow{2}{*}{$210$} & Raw & $-0.211(14)$ & $-0.212(15)$ & $<1\sigma$ \\
 & & & DRC & $-0.212(14)$ & $-0.211(20)$ & $<1\sigma$ \\
\hline
\multirow{4}{*}{$(0,\frac{3}{2})$} &\multirow{4}{*}{$\Xi_{cc} \pi \to \Xi_{cc} \pi$ } & \multirow{2}{*}{$300$} & Raw & $-0.166(11)$ & $-0.111(24)$ & $2.1\sigma$ \\
 & & & DRC & $-0.140(14)$ & $-0.110(22)$ & $1.1\sigma$ \\
\cline{3-7}
 & & \multirow{2}{*}{$210$} & Raw & $-0.149(21)$ & $-0.137(26)$ & $<1\sigma$ \\
 & & & DRC & $-0.143(24)$ & $-0.140(27)$ & $<1\sigma$ \\
\hline
\end{tabular}
\caption{Comparison of the scattering lengths $a_0$ determined from LO and NLO ERE fits, with differences quantified in standard deviations ($\sigma$).}
\label{tab:ERE.FitRslts.P2vsP4}
\end{center}
\end{table}

\begin{figure}[ht] 
\begin{center}
  \begin{subfigure}[b]{0.45\linewidth}
    \centering
    \includegraphics[width=1.0\linewidth]{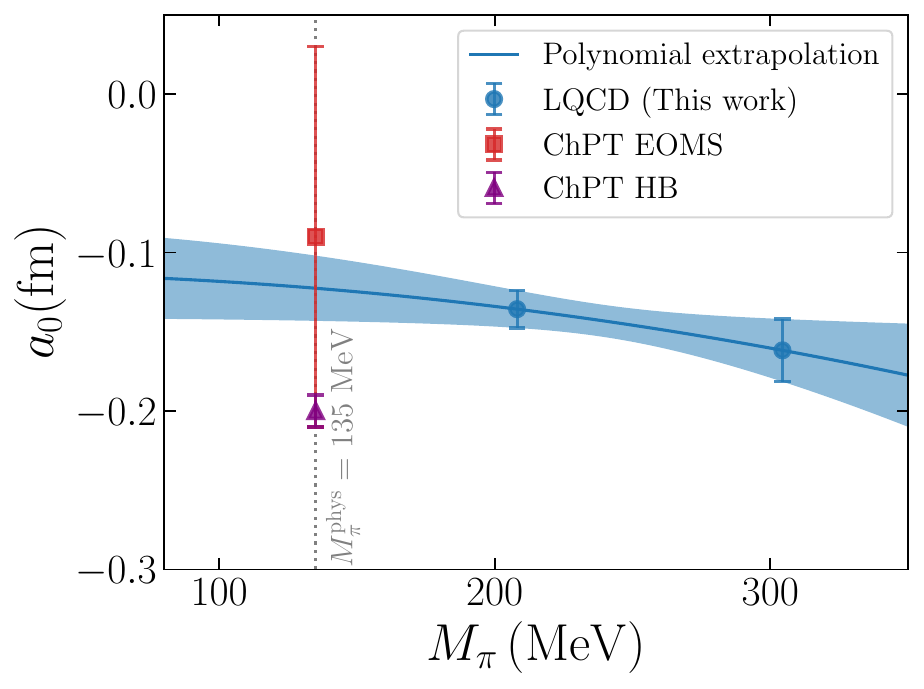} 
    \caption{$\Omega_{cc}\bar{K}^{(-2,\frac{1}{2})}$} 
    \label{fig:ScatLen.extrapolation:a} 
  \end{subfigure}
  \begin{subfigure}[b]{0.45\linewidth}
    \centering
    \includegraphics[width=1.0\linewidth]{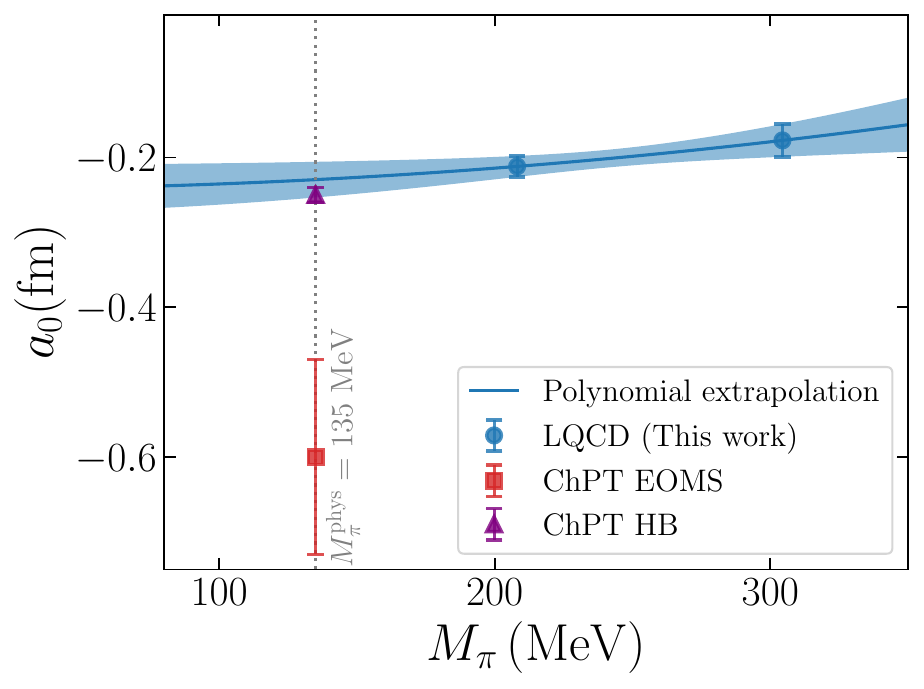} 
    \caption{$\Xi_{cc}K^{(1,1)}$} 
    \label{fig:ScatLen.extrapolation:b} 
  \end{subfigure} 
  \begin{subfigure}[b]{0.45\linewidth}
    \centering
    \includegraphics[width=1.0\linewidth]{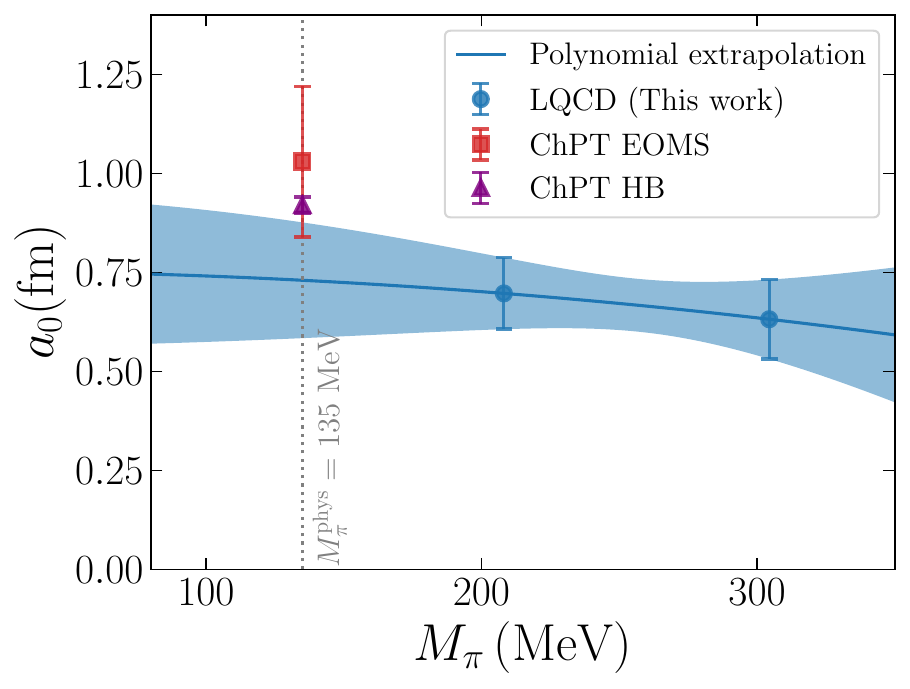} 
    \caption{$\Xi_{cc}K^{(1,0)}$} 
    \label{fig:ScatLen.extrapolation:c} 
  \end{subfigure}
  \begin{subfigure}[b]{0.45\linewidth}
    \centering
    \includegraphics[width=1.0\linewidth]{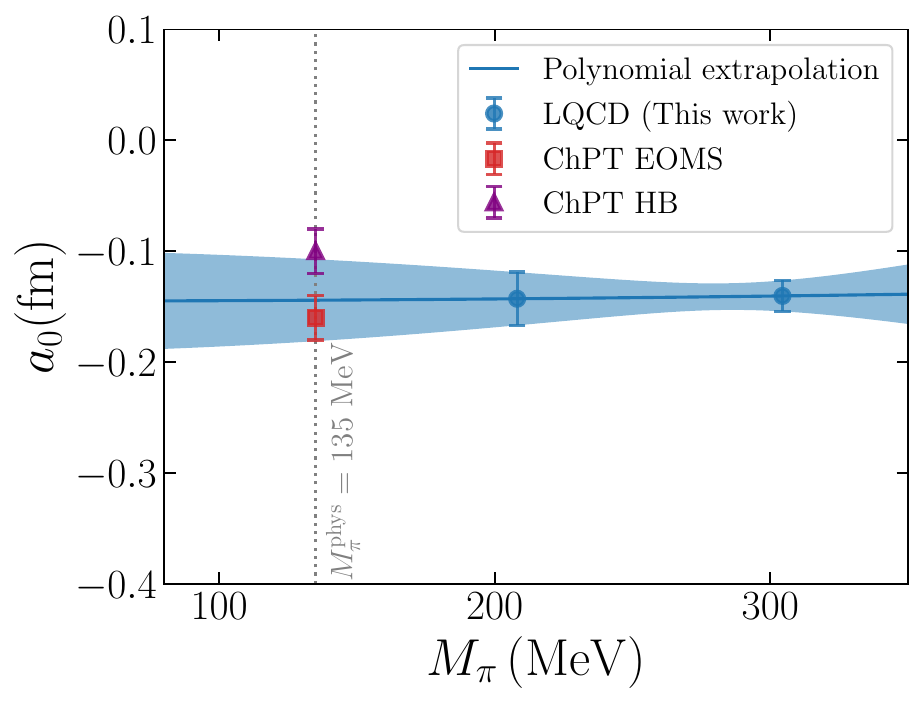} 
    \caption{$\Xi_{cc}\pi^{(0,\frac{3}{2})}$} 
    \label{fig:ScatLen.extrapolation:d} 
  \end{subfigure} 
\caption{Extrapolation of the $S$-wave scattering lengths $a_0$ to the physical pion mass. The results of $a_0$ labeled by \lq\lq DRC" from table~\ref{tab:ERE.FitRslts.UpToP2}\ are adopted in the extrapolation. The EOMS- and HB-ChPT predictions, taken from refs.~\cite{Liang:2023scp,Meng:2018zbl} respectively, are also shown for easy comparison.}
\label{fig:ScatLen.extrapolation}
\end{center}
\end{figure}

In the above, the $S$-wave scattering lengths for the four distinct channels have been determined at two unphysical pion masses, $\sim$ 300 and 210 MeV. To extrapolate them to the physical pion mass $M_{\pi}^{\mathrm{phy}}=135$ MeV, we employ a rank-one polynomial in $M_\pi^2$ inspired by BChPT as follows:
\begin{equation}
a_0(M_{\pi}^2)=\alpha_0 + \alpha_1M_{\pi}^2\ .\label{eq.chi.extrap}
\end{equation}
This linear dependence corresponds to the $\mathcal{O}(p^2)$ result in BChPT. Given that only two pion masses are available, this extrapolation serves as an initial estimate. A more comprehensive extrapolation using the chiral expression from ref.~\cite{Liang:2023scp} is only feasible once more data points are available from lattice simulations at different pion masses in the future. In figure~\ref{fig:ScatLen.extrapolation}, the blue points represent lattice results at two simulated pion masses, while the blue line and band represent the linear extrapolation with its 1$\sigma$ statistical uncertainty. For comparison, the ChPT results obtained within the EOMS scheme~\cite{Liang:2023scp} and HB formalism~\cite{Meng:2018zbl} are displayed as well. 

Our final predictions of scattering lengths at the physical pion mass, with statistical and two systematic uncertainties listed sequentially, are:
\begin{align}
    a_0^{\mathrm{phy}}(\Xi_{cc}K^{(1,1)}) &= -0.230 (24)_{\rm sta} (5)_{{\rm sys}_1} (6)_{{\rm sys}_2} \ \mathrm{fm} = -0.230(25) \ \mathrm{fm}\ ,  \\
    a_0^{\mathrm{phy}}(\Xi_{cc}K^{(1,0)}) &= 0.730 (146)_{\rm sta} (13)_{{\rm sys}_1} \ \mathrm{fm}=0.730 (147)\ \mathrm{fm}\ ,  \\
    a_0^{\mathrm{phy}}(\Xi_{cc}\pi^{(0,3/2)}) &= -0.144 (37)_{\rm sta} (3)_{{\rm sys}_1} (11)_{{\rm sys}_2}\ \mathrm{fm} =-0.144 (39) \ \mathrm{fm}\ ,  \\
    a_0^{\mathrm{phy}}(\Omega_{cc}\bar{K}^{(-2,1/2)}) &= -0.123 (21)_{\rm sta} (14)_{{\rm sys}_1} (2)_{{\rm sys}_2} \ \mathrm{fm}=-0.123(25)\ \mathrm{fm}\ . 
\end{align}
The central values of the extrapolated results are obtained by performing the linear fit~\eqref{eq.chi.extrap} to the scattering lengths from the LO ERE analysis with DRC applied. The first uncertainty quoted is the statistical error (sta) propagated from the $1$-$\sigma$ uncertainties of the fit parameters. We estimate two dominant sources of systematic uncertainty, both originating from the choice of input values for the extrapolation. The first systematic uncertainty (sys$_1$) is associated with the DRC, which we quantify as the difference between the central extrapolated value and that obtained using the Raw dataset. The second systematic uncertainty (sys$_2$) stems from the truncation of the ERE, estimated as the difference when using the results from the NLO fits as input for the extrapolation. It is worth noting that this error budget represents a rough estimation based on current data. To achieve a robust control over the systematic uncertainties, calculations at more pion masses and lattice spacings are mandatory. Note that for the $\Xi_{cc}K^{(1,0)}$ channel, the systematic uncertainty from the ERE truncation is not quoted, as an NLO fit was not feasible due to the limited number of data points.

Overall, our results are in good agreement with ChPT predictions at the physical point. Specifically, the scattering lengths of the $\Xi_{cc}K^{(1,0)}$ and $\Xi_{cc}\pi^{(0,\frac{3}{2})}$ channels are consistent with both the EOMS and HB determinations within 1.5$\sigma$. In addition, our results favor the EOMS predictions in the $\Omega_{cc}\bar{K}^{(-2,\frac{1}{2})}$ channel but align more closely with the HB predictions in the $\Xi_{cc}K^{(1,1)}$ channel. A crucial caveat for these ChPT predictions is that they rely on the LECs, whose values are estimated from $D\phi$ interactions via HDA symmetry. Note also that for the channels with large scattering lengths, it is actually necessary to refine the chiral amplitudes by performing unitarization, as done in refs.~\cite{Guo:2017vcf,Yan:2018zdt}.

\subsection{Pole analysis of the scattering amplitude}
In $S$-matrix theory, poles of the elastic scattering amplitude on the real axis of the $s$-plane below the threshold are identified as bound states (the first Riemann sheet, Im\,$p>0$) or virtual states (the second Riemann sheet, Im\,$p<0$). Equivalently, in the $p$-plane, bound and virtual states reside on the positive and negative imaginary axes, respectively, which can be readily inferred from $p={\sqrt{(s-s_+)(s-s_-)}}/{(2\sqrt{s})}$ with $s_\pm=(M_{B_{cc}}\pm M_{\phi})^2$. Therefore, in view of eq.~(\ref{eq:scat.amp}), an efficient way to locate these poles is to identify the intersections between the $p\cot\delta_0$ and $ip = \pm |p|$ curves in the range of $p^2 < 0$. To that end, we plot $ip=\pm |p|$ in addition to $p\cot\delta_0$ in figure~\ref{fig:ERE.shifted}, which are represented by the gray curves. The intersections in the upper and lower half-plane correspond to virtual and bound states, respectively.

Interestingly, the $\Omega_{cc}\bar{K}^{(-2,\frac{1}{2})}$ and $\Xi_{cc}K^{(1,1)}$ channels, previously regarded as repulsive due to upward energy shifts in~\cref{sec.4pt.cf.el}, exhibit bound-state poles in LO ERE fits. The appearance of these unexpected poles is attributed to the LO-truncated ERE and sparse data. On the one hand, the pole positions shift significantly or disappear entirely in the NLO fits, as can be seen from the divergent yellow uncertainty bands in figure~\ref{fig:ERE.shifted}, demonstrating the instability of the LO results. On the other hand, the poles are located in the range $p^2<-0.20\ \mathrm{GeV}^2$, which is far beyond the region sampled by our finite-volume energy levels. The resultant existence of those poles is unreliable due to the lack of data to constrain the scattering amplitude in the $p^2<-0.20\ \mathrm{GeV}^2$ range. Therefore, we refrain from interpreting those poles as any possible physical states in reality. The $\Xi_{cc}\pi^{(0,\frac{3}{2})}$ channel is repulsive, and no poles are found, as can be seen from \cref{fig:ERE.shifted.d}.

The interaction in the $\Xi_{cc}K^{(1,0)}$ channel is attractive, manifesting as negative energy shifts in \cref{fig:energy.levels}. From the perspective of HDA symmetry, this channel corresponds to the $I=0$ channel of $D\bar{K}$ scattering. Previous studies have reported a virtual state in the corresponding $D\bar{K}$ channel~\cite{Guo:2015dha,Cheung:2020mql}. Motivated by this, we investigate the pole position in the $\Xi_{cc}K^{(1,0)}$ system to explore the potential existence of a counterpart state. A near-threshold pole at $p^2\approx-0.03\ \mathrm{GeV}^2$ is discovered in the upper half-plane for both $M_\pi\sim 210$ and $300$~MeV, as shown in \cref{fig:ERE.shifted.c}. This pole suggests the existence of a possible virtual state and indicates that the strength of the attraction is too weak to form a bound state. However, the stability of this pole against ERE truncation remains unevaluated due to insufficient data for performing NLO fits.

\section{Summary and outlook\label{sec.sum}}

We have presented the first lattice QCD study of $S$-wave interactions between doubly charmed baryons $B_{cc}\in\{\Xi_{cc}, \Omega_{cc}\}$ and Nambu-Goldstone bosons $\phi\in\{\pi, K
\}$ at two pion masses $M_{\pi}\sim 210$ and $300$ MeV using four $2+1$ flavor full-QCD ensembles provided by the CLQCD collaboration. Finite-volume energy levels for the four single channels, $\Omega_{cc}\bar{K}^{(-2,\frac{1}{2})}$, $\Xi_{cc}{K}^{(1,1)}$,  $\Xi_{cc}{K}^{(1,0)}$ and $\Xi_{cc}{K}^{(0,\frac{3}{2})}$, are obtained. We find that the $\Xi_{cc}{K}^{(1,0)}$ channel is attractive due to negative energy shifts from the non-interacting thresholds, while the other channels are repulsive. It is worth noting that the same interacting information was also identified for the four single-channel $D\phi$ interactions in ref.~\cite{Liu:2012zya}, subject to the degeneracy between the $D\phi$ and $B_{cc}\phi$ systems under HDA symmetry. The phase shifts near thresholds are extracted by utilizing L\"uscher's formula; consequently, the scattering lengths and effective ranges in the ERE are determined. Our results for the scattering lengths show robustness against the ERE truncation order and agree with previous BChPT predictions after extrapolation to the physical point. Interestingly, pole analysis near the threshold implies a $J^P=1/2^-$ virtual state in the $\Xi_{cc}{K}^{(1,0)}$ channel, located at $p^2\approx-0.03\ \mathrm{GeV}^2$. 
This work provides first-principle lattice-QCD inputs for theoretical investigations of double-heavy baryon spectroscopy, complementing ongoing experimental efforts.
The present single-channel studies also pave the way for future exploration of the coupled-channel $B_{cc}\phi$ interactions. 

\acknowledgments
We would like to thank Ying~Chen, Feng-Kun~Guo, Qu-Zhi~Li, Chuan~Liu, Peng~Sun, Xiao-Nu~Xiong and Yi-Bo~Yang for helpful discussions. We thank the CLQCD collaboration for providing us their gauge configurations with dynamical fermions~\cite{CLQCD:2023sdb,CLQCD:2024yyn}, which are generated on the Southern Nuclear Science Computing Center (SNSC), HPC Cluster of ITP-CAS, and the Siyuan-1 cluster supported by the Center for High Performance Computing at Shanghai Jiao Tong University. This work is supported by Science Research Project of Hebei Education Department under Contract No.~QN2025063; National Natural Science Foundations of China (NSFC) under Contract Nos. 12275076, 11905258, 12335002, 12175279, 12293060, 12293061; by the Science Fund for Distinguished Young Scholars of Hunan Province under Grant No.~2024JJ2007; by Hebei Natural Science Foundation under Grant No.~A2025205018; by the Science Foundation of Hebei Normal University with Contract No.~L2025B09.

\appendix

\section{Extraction of energies\label{sec.res.cf}}
\begin{figure}[htb]
\centering
\includegraphics[width=0.78\linewidth]{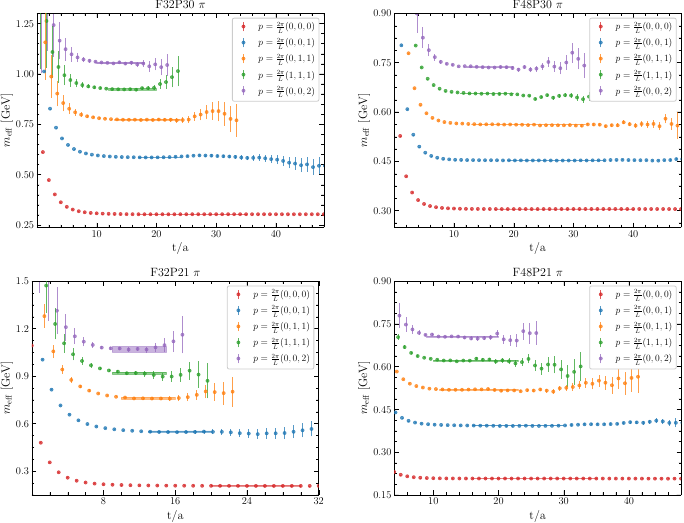} 
\caption{The pion effective mass plots for four ensembles. Each panel displays the results for the five lowest momenta, $\mathbf{p}$ = (0, 0, 0), (0, 0, 1), (0, 1, 1), (1, 1, 1), and (0, 0, 2), in units of $2\pi/L$. The horizontal lines with bands indicate the fitted masses and the fitting ranges.} 
\label{fig:pion.meff}
\end{figure}

\begin{figure}[htb]
\centering
\includegraphics[width=0.78\linewidth]{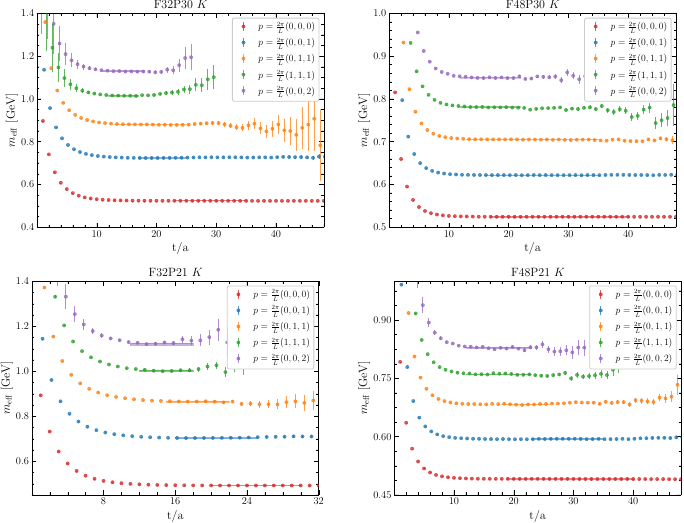} 
\caption{The kaon effective mass plots for four ensembles. The description is the same as in figure~\ref{fig:pion.meff}.} 
\label{fig:kaon.meff}
\end{figure}

\begin{figure}[htb]
\centering
\includegraphics[width=0.78\linewidth]{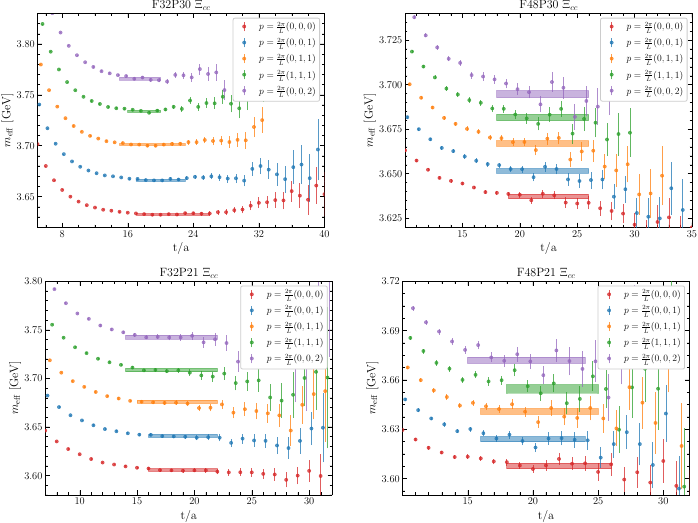} 
\caption{The $\Xi_{cc}$ effective mass plots for four ensembles. The description is the same as in figure~\ref{fig:pion.meff}.} 
\label{fig:Xicc.meff}
\end{figure}

\begin{figure}[htb]
\centering
\includegraphics[width=0.78\linewidth]{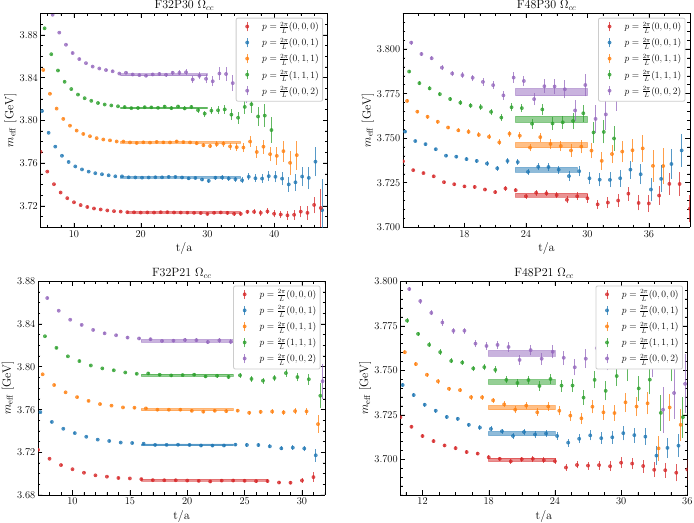} 
\caption{The $\Omega_{cc}$ effective mass plots for four ensembles. The description is the same as in figure~\ref{fig:pion.meff}.} 
\label{fig:Omegacc.meff}
\end{figure}

\begin{figure}[htb]
\centering
\includegraphics[width=0.99\linewidth]{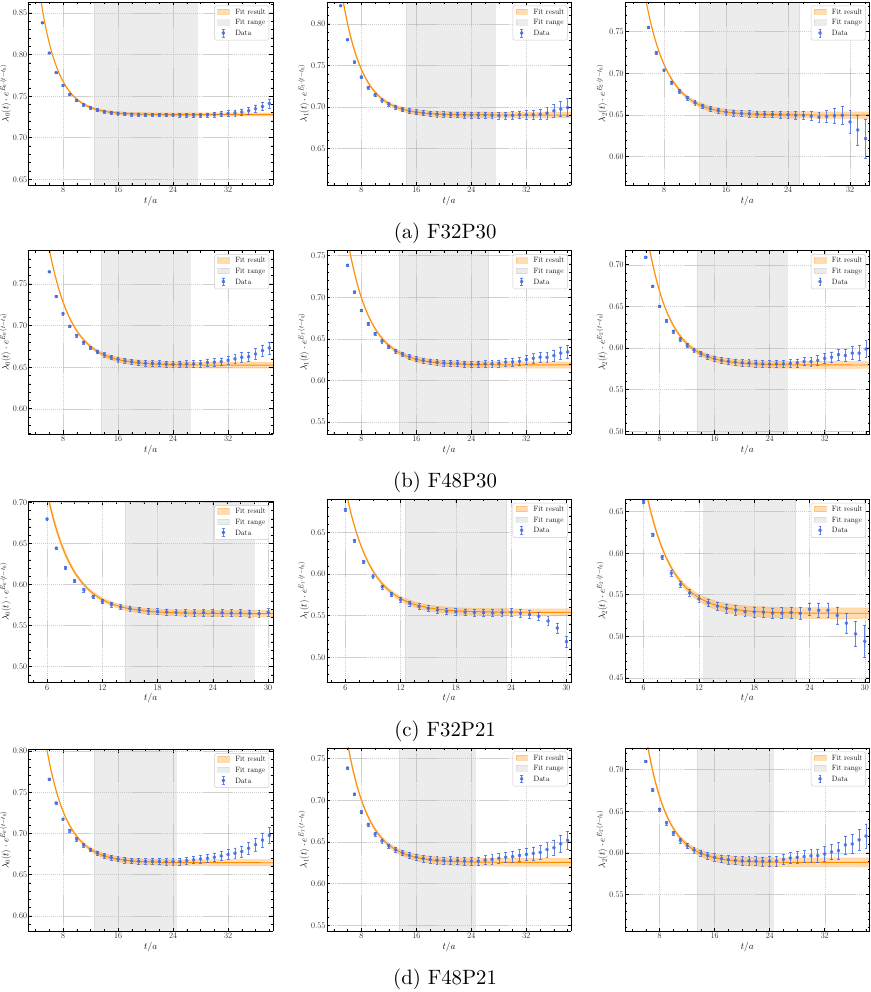} 
\caption{Fits of the GEVP eigenvalues $\lambda_n$ for the $\Omega_{cc}\bar{K}^{(-2,1/2)}$ system. Plotted is the rescaled quantity $\lambda_n(t)e^{E_n(t-t_0)}$, which is expected to be constant at large time slices. The data points are shown in blue, while the yellow lines and bands represent the fit results with their statistical errors. The gray shaded areas indicate the temporal range used for the fits. Panels (a)-(d) show results on four different ensembles. Within each panel, the plots from left to right correspond to the three lowest momenta, corresponding to $|\mathbf{p}| = \{0, 1, \sqrt{2}\}$ (in units of $2\pi/L$).}
\label{fig:OmegaccKbar.meff}
\end{figure}

\begin{figure}[htb]
\centering
\includegraphics[width=0.99\linewidth]{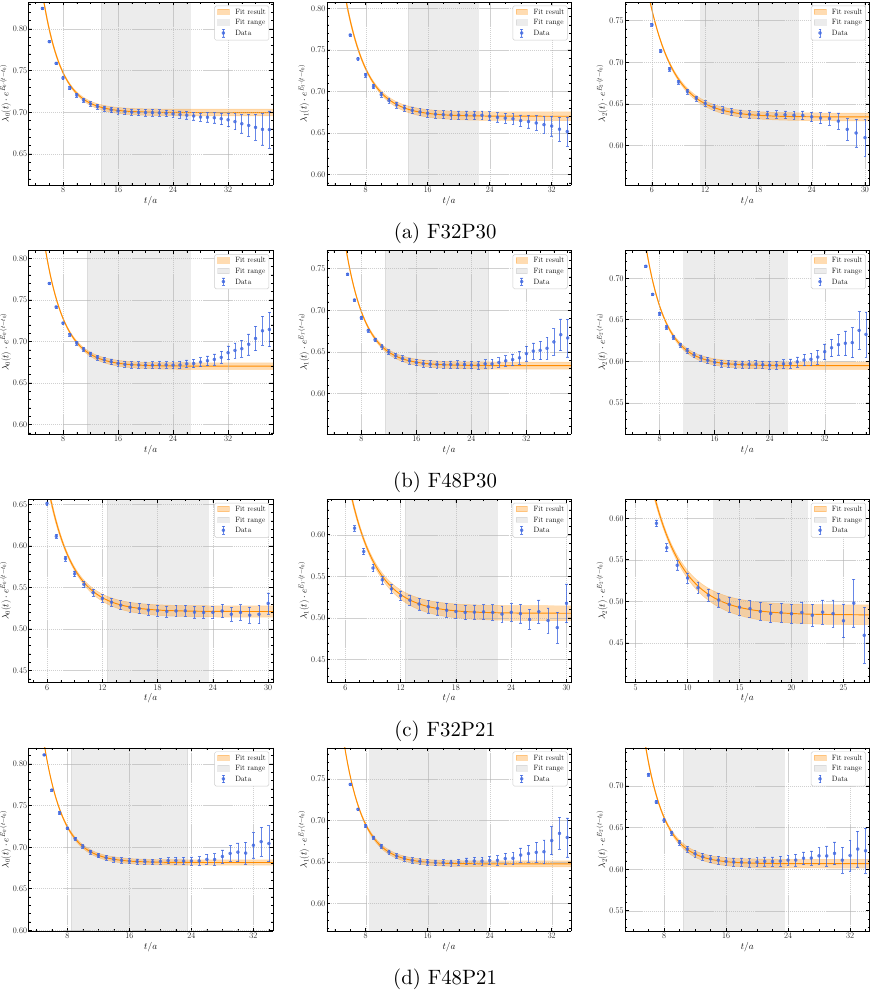} 
\caption{Fits of the GEVP eigenvalues $\lambda_n$ for the $\Xi_{cc}K^{(1,0)}$ system on four ensembles. The description is the same as in figure~\ref{fig:OmegaccKbar.meff}.} 
\label{fig:XiccK10.meff}
\end{figure}

\begin{figure}[htb]
\centering
\includegraphics[width=0.99\linewidth]{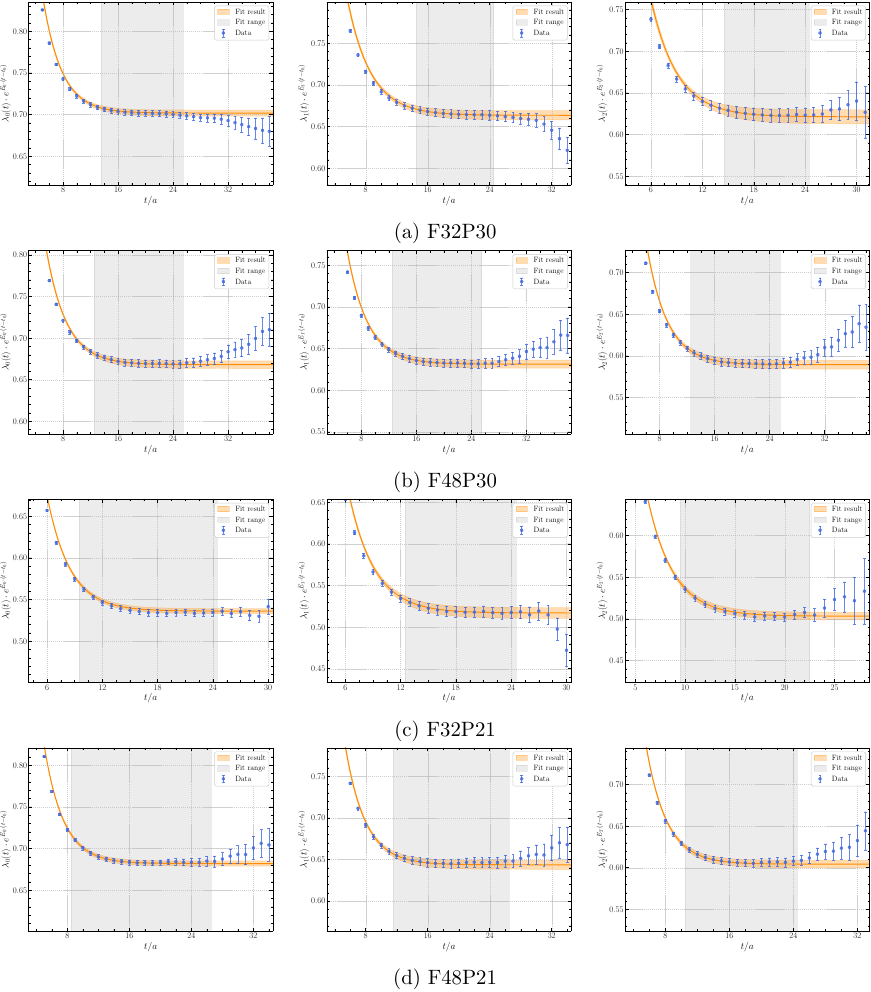} 
\caption{Fits of the GEVP eigenvalues $\lambda_n$ for the $\Xi_{cc}K^{(1,1)}$ system on four ensembles. The description is the same as in figure~\ref{fig:OmegaccKbar.meff}.} 
\label{fig:XiccK11.meff}
\end{figure}

\begin{figure}[htb]
\centering
\includegraphics[width=0.99\linewidth]{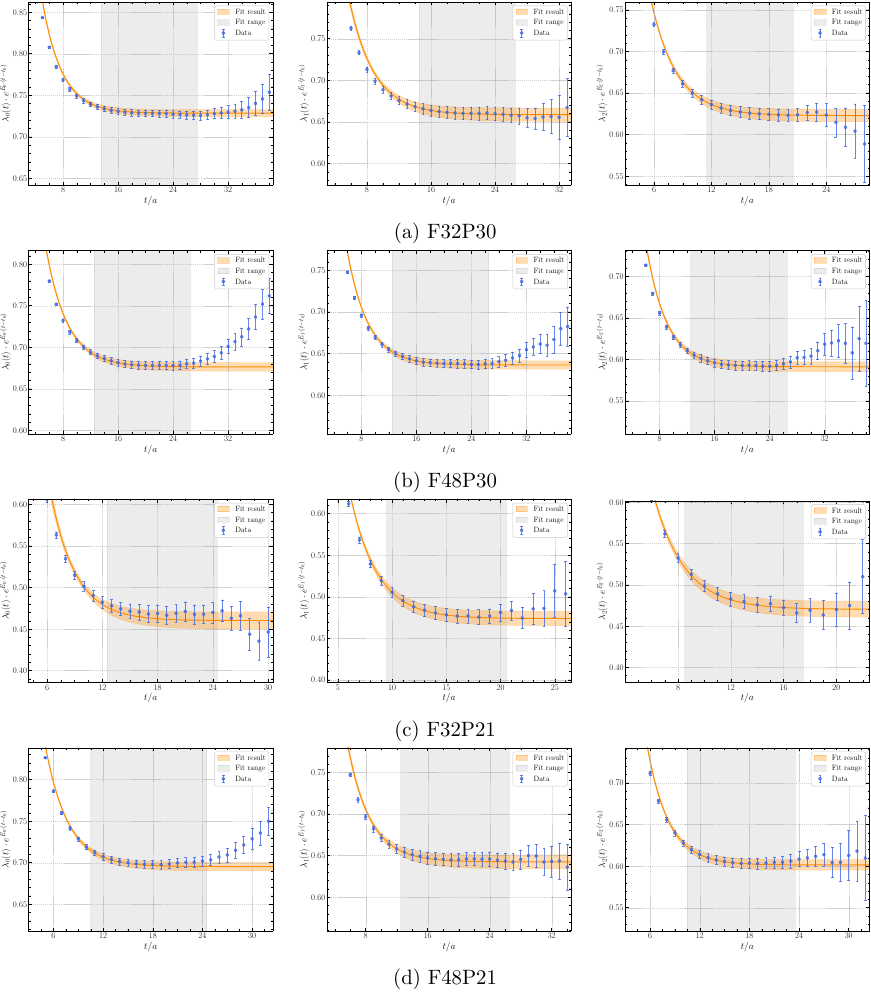} 
\caption{Fits of the GEVP eigenvalues $\lambda_n$ for the $\Xi_{cc}\pi^{(0,3/2)}$ system on four ensembles. The description is the same as in figure~\ref{fig:OmegaccKbar.meff}.} 
\label{fig:XiccPi.meff}
\end{figure}

In this section, we present the plots used to determine the masses of single hadrons and to illustrate the quality of the energy level fits for two-hadron systems. The effective masses for the $\pi$, $K$, $\Xi_{cc}$ and $\Omega_{cc}$ are shown in figures~\ref{fig:pion.meff}, \ref{fig:kaon.meff}, \ref{fig:Xicc.meff} and \ref{fig:Omegacc.meff}, respectively. In these figures, the horizontal lines with bands indicate the fitted mass values and the corresponding temporal ranges used for the fits.
For the two-hadron systems, figures~\ref{fig:OmegaccKbar.meff}, \ref{fig:XiccK10.meff}, \ref{fig:XiccK11.meff} and~\ref{fig:XiccPi.meff} present the results from the two-exponential fits to the GEVP eigenvalues. To better visualize the fit quality, these figures plot the rescaled eigenvalues, which are expected to approach a constant at large time slices. The close agreement between the data points and the fit curves shown in the plots confirms the reliability of our extracted energy levels.



\clearpage
\section{Details of the interpolating operators}\label{appendix.otherchannels}
The single-hadron interpolating operators for the Goldstone bosons, $\pi^+$, $K^+$ and $K^0$, are given by
\begin{align}
\mathcal{O}_{\pi^+}(x)&=\bar{d}(x)^a_{\alpha} (\gamma_5)_{\alpha \beta} u(x)^a_{\beta}\ ,\notag\\
\mathcal{O}_{K^+}(x)&=\bar{s}(x)^a_{\alpha } (\gamma_5)_{\alpha \beta} u(x)^a_{\beta}\ ,\notag\\
\mathcal{O}_{K^0}(x)&=\bar{s}(x)^a_{\alpha} (\gamma_5)_{\alpha \beta} d(x)^a_{\beta}\ ,\label{eq.interpolator.meson}
\end{align}
where the Greek and Latin letters denote the Dirac and color indices, respectively; and summation over repeated indices is implied. The interpolators for $\pi^-$, $K^-$ and $\bar{K}^0$ can be readily obtained by performing charge conjugation. For the ground states of doubly charmed baryons $\Xi_{cc}^{++,+}$ and $\Omega_{cc}^+$ with $J^P=(1/2)^+$, the interpolators read:
\begin{align}
\mathcal{O}_{\Xi_{cc}^{++}}(x)&=\epsilon^{ijk}P_{+}[c^{i T}(x)C\gamma_5 u^j(x)] c^k(x)\ , \notag \\
\mathcal{O}_{\Xi_{cc}^{+\,\,}}(x)&=\epsilon^{ijk}
P_{+}[c^{i T}(x) C\gamma_5 d^j(x)] c^k(x)\ ,\notag
 \\
\mathcal{O}_{\Omega_{cc}^{+\,\,}}(x)&=\epsilon^{ijk}
P_{+}[c^{i T}(x) C\gamma_5 s^j(x)] c^k(x)\ ,\label{eq.interpolator.baryon}
\end{align}
where $C=i\gamma_2\gamma_0$ is the charge conjugation matrix and the positive parity projector is defined by $P_+=(1+\gamma_0)/2$. For the two-particle operators, the coefficients $C_{\alpha, \mathbf{p}_{\mathbf{1}}, \mathbf{p}_{\mathbf{2}}}$ in eq.~\eqref{eq.our.twoparticle.ops} are listed in Table~\ref{tab:two_op_coefs}. 
\begin{table}[ht]
\centering
\begin{tabular}{c|c|c|c|c}
\hline
& $\alpha$ & $\mathbf{p}_{\mathbf{1}}$ & $\mathbf{p}_{\mathbf{2}}$ & $C_{\alpha, \mathbf{p}_{\mathbf{1}}, \mathbf{p}_{\mathbf{2}}}$ \\
\hline 
$p=0$ & $1$ & $(0,0,0)$ & $(0,0,0)$ & $1$ \\
\hline 
\multirow{6}{*}{$p=1$} & 1 & $(-1,0,0)$ & $(1,0,0)$ & $1$ \\
\cline { 2 - 5 } & $1$ & $(1,0,0)$ & $(-1,0,0)$ & $1$ \\
\cline { 2 - 5 } & $1$ & $(0,-1,0)$ & $(0,1,0)$ & $1$ \\
\cline { 2 - 5 } & $1$ & $(0,1,0)$ & $(0,-1,0)$ & $1$ \\
\cline { 2 - 5 } & $1$ & $(0,0,-1)$ & $(0,0,1)$ & $1$ \\
\cline { 2 - 5 } & $1$ & $(0,0,1)$ & $(0,0,-1)$ & $1$ \\
\hline
\multirow{11}{*}{$p=\sqrt{2}$} & $1$ & $(-1,-1,0)$ & $(1,1,0)$ & $1$ \\
\cline { 2 - 5 } & $1$ & $(1,1,0)$ & $(-1,-1,0)$ & $1$ \\
\cline { 2 - 5 } & $1$ & $(-1,0,-1)$ & $(1,0,1)$ & $1$ \\
\cline { 2 - 5 } & $1$ & $(1,0,1)$ & $(-1,0,-1)$ & $1$ \\
\cline { 2 - 5 } & $1$ & $(0,-1,-1)$ & $(0,1,1)$ & $1$ \\
\cline { 2 - 5 } & $1$ & $(0,1,1)$ & $(0,-1,-1)$ & $1$ \\
\cline { 2 - 5 } & $1$ & $(-1,1,0)$ & $(1,-1,0)$ & $1$ \\
\cline { 2 - 5 } & $1$ & $(1,-1,0)$ & $(-1,1,0)$ & $1$ \\
\cline { 2 - 5 } & $1$ & $(-1,0,1)$ & $(1,0,-1)$ & $1$ \\
\cline { 2 - 5 } & $1$ & $(1,0,-1)$ & $(-1,0,1)$ & $1$ \\
\cline { 2 - 5 } & $1$ & $(0,1,-1)$ & $(0,-1,1)$ & $1$ \\
\cline { 2 - 5 } & $1$ & $(0,-1,1)$ & $(0,1,-1)$ & $1$ \\
\hline
\end{tabular}
\caption{Values of the coefficients $C_{\alpha, \mathbf{p}_{\mathbf{1}}, \mathbf{p}_{\mathbf{2}}}$ in the two-particle operators~\eqref{eq.our.twoparticle.ops} for the momenta $p^2=0,1,2$, in units of $(2\pi/L)^2$. }
\label{tab:two_op_coefs}
\end{table}

Furthermore, as mentioned in section~\ref{sec.4pt.cf.el}, we must consider some channels that may be encountered in our analysis of the two-particle system. The operators of $D^+$, $D^0$, $D_s^+$, $\Lambda_{c}^{+}$, $\Sigma_{c}^{++}$, and $\Omega_{c}^{0}$ are needed, which are constructed as
\begin{align}
\mathcal{O}_{D^+}(x)&=\bar{d}(x)^a_{\alpha} (\gamma_5)_{\alpha \beta} c(x)^a_{\beta},  \notag \\
\mathcal{O}_{D^0}(x)&=\bar{u}(x)^a_{\alpha} (\gamma_5)_{\alpha \beta} c(x)^a_{\beta},  \notag \\
\mathcal{O}_{D_s^+}(x)&=\bar{s}(x)^a_{\alpha} (\gamma_5)_{\alpha \beta} c(x)^a_{\beta} ,  \notag\\
\mathcal{O}_{\Lambda_{c}^{+}(udc)}(x)&=\epsilon^{ijk}P_{+}[Q_u^{i T}(x)C\gamma_5 q_d^j(x)] Q_c^k(x)\ , \notag\\
\mathcal{O}_{\Sigma_{c}^{++}(uuc)}(x)&=\epsilon^{ijk}P_{+}[Q_u^{i T}(x)C\gamma_5 q_c^j(x)] Q_u^k(x)\ , \notag\\
\mathcal{O}_{\Omega_{c}^{0}(ssc)}(x)&=\epsilon^{ijk}P_{+}[Q_s^{i T}(x)C\gamma_5 q_c^j(x)] Q_s^k(x)\ .\label{eq.Bc.D.interpolators}
\end{align}
We perform simulations of the two-point correlation functions with the above interpolating fields. The corresponding masses are obtained and are listed in table~\ref{tab:meff.others}. The thresholds of $B_c D$ channels are illustrated in figure~\ref{fig:energy.levels}. 

\begin{table}[htb]
\begin{center}
\begin{tabular}{c|c|c|c|c|c}
\hline
      &       $D$&     $D_s$& $\Lambda_c$& $\Sigma_c$& $\Omega_c$
\\
\hline 
F48P30& $1.8923(4)$  & $1.9706(2)$ & $2.3488(9)$& $2.4943(19)$& $2.6703(11)$
\\
\hline 
F48P21& $1.8736(4)$& $1.9595(3)$&  $2.2842(18)$& $2.4438(18)$&$2.6420(10)$
\\
\hline 
\end{tabular}
\caption{Masses of $D$, $D_s$, $\Lambda_c$, $\Sigma_c$, and $\Omega_c$ in units of GeV.}
\label{tab:meff.others}
\end{center}
\end{table}


\section{Wick contractions\label{sec.con}}

\begin{figure}[htb]
\centering
\includegraphics[width=0.99\linewidth]{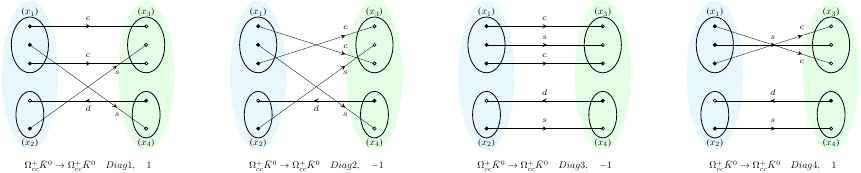} 
\caption{Sketch of the quark contraction for $\Omega_{cc} \bar{K} \to \Omega_{cc} \bar{K}$ with $(S,I)=(-2,1/2)$.} 
\label{fig:OmgccKbar_m2_0dot5_contraction}
\end{figure}

\begin{figure}[htb]
\centering
\includegraphics[width=0.99\linewidth]{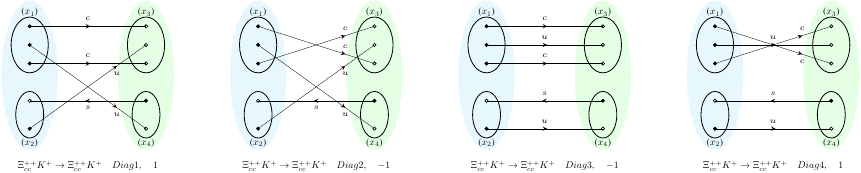} 
\caption{Sketch of the quark contraction for $\Xi_{cc} K \to \Xi_{cc} K$ with $(S,I)=(1,1)$.} 
\label{fig:XiccK_1_1_contraction}
\end{figure}

\begin{figure}[htb]
\centering
\includegraphics[width=0.99\linewidth]{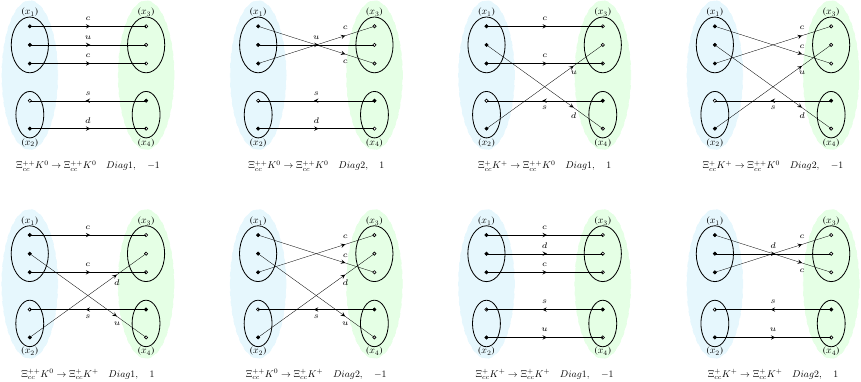} 
\caption{Sketch of the quark contraction for $\Xi_{cc} K \to \Xi_{cc} K$ with $(S,I)=(1,0)$.
} 
\label{fig:XiccK_1_0_contraction}
\end{figure}

\begin{figure}[htb]
\centering
\includegraphics[width=0.99\linewidth]{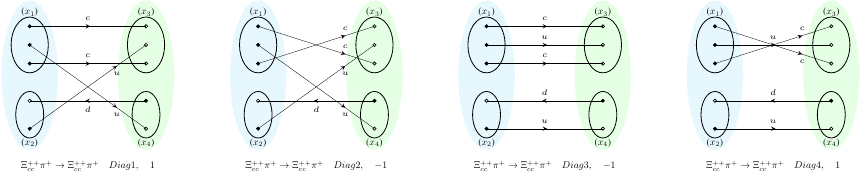} 
\caption{Sketch of the quark contraction for $\Xi_{cc} \pi \to \Xi_{cc} \pi$ with $(S,I)=(0,\frac{3}{2})$.} 
\label{fig:XiccPi_contraction}
\end{figure}

For completeness, we show the quark contraction diagrams for all four channels that are considered in our numerical computation in \cref{fig:OmgccKbar_m2_0dot5_contraction,fig:XiccK_1_1_contraction,fig:XiccK_1_0_contraction,fig:XiccPi_contraction}.

\clearpage


\bibliographystyle{JHEP}
\bibliography{biblio.bib}

\end{document}